
\newcount\mgnf\newcount\tipi\newcount\tipoformule
\newcount\aux\newcount\driver\newcount\cind\newcount\bz
\newcount\tipobib\newcount\stile

\newdimen\stdindent\newdimen\bibskip
\newdimen\maxit\maxit=0pt

\stile=0         
\tipobib=1       
\bz=0            
\cind=0          
\mgnf=1          
\tipoformule=0   
\aux=0           


\ifnum\mgnf=0
   \magnification=\magstep0 
   \hsize=17truecm\vsize=24truecm\hoffset=-1cm
   \parindent=4.pt\stdindent=\parindent\fi
\ifnum\mgnf=1
   \magnification=\magstep1\hoffset=-0.7truecm
   \voffset=-1.0truecm\hsize=17truecm\vsize=24.truecm
   \baselineskip=14pt plus0.1pt minus0.1pt \parindent=6pt
   \lineskip=4pt\lineskiplimit=0.1pt      \parskip=0.1pt plus1pt
   \stdindent=\parindent\fi


\def\fine#1{}
\def\draft#1{\bz=1\ifnum\mgnf=1\baselineskip=22pt \else\baselineskip=16pt\fi
   \ifnum\stile=0\headline={\hfill DRAFT #1}\fi\raggedbottom
    \def\gnuins ##1 ##2 ##3{\gnuinsf {##1} {##2} {##3}}
    \def\gnuin ##1 ##2 ##3 ##4 ##5 ##6{\gnuinf {##1} {##2} {##3} 
		{##4} {##5} {##6}}
      \def\fine ##1{\vfill\eject\centerline{\bf Figures' captions}
		\def\geq(####1){}
                \unvbox150\vfill\eject
                \centerline{FIGURES}\unvbox149 ##1}}

\def\large{\bz=0\ifnum\mgnf=1\baselineskip=22pt \else\baselineskip=16pt\fi
   \raggedbottom
    \def\gnuins ##1 ##2 ##3{\gnuinsf {##1} {##2} {##3}}
    \def\gnuin ##1 ##2 ##3 ##4 ##5 ##6{\gnuinf {##1} {##2} {##3} 
		{##4} {##5} {##6}}
      \def\fine ##1{\vfill\eject\centerline{\bf Figures' captions}
		\def\geq(####1){}
                \unvbox150\vfill\eject
                \centerline{FIGURES}\unvbox149 ##1}}


\newcount\prau

\def\titolo#1{\setbox197\vbox{ 
\leftskip=0pt plus16em \rightskip=\leftskip
\spaceskip=.3333em \xspaceskip=.5em \parfillskip=0pt
\pretolerance=9999  \tolerance=9999
\hyphenpenalty=9999 \exhyphenpenalty=9999
\ftitolo #1}}
\def\abstract#1{\setbox198\vbox{
     \centerline{\vbox{\advance\hsize by -2cm \parindent=0pt\it Abstact: #1}}}}
\def\parole#1{\setbox195\hbox{
     \centerline{\vbox{\advance\hsize by -2cm \parindent=0pt Keywords: #1.}}}}
\def\autore#1#2{\setbox199\hbox{\unhbox199\ifnum\prau=0 #1%
\else, #1\fi\global\advance\prau by 1$^{\simbau}$}
     \setbox196\vbox {\advance\hsize by -\parindent\copy196$^{\simbau}${#2}}}
\def\prima{\unvbox197\vskip1truecm\centerline{\unhbox199}
     \footnote{}{\unvbox196}\vskip1truecm\unvbox198\vskip1truecm\copy195}
\def\simbau{\ifcase\prau
	\or \dagger \or \ddagger \or * \or \star \or \bullet\fi}


       \let\d=\delta 
      \let\l=\lambda
                   
 \let\t=\tau

\let\P=\Pi            \let\O=\Omega


{\count255=\time\divide\count255 by 60 \xdef\oramin{\number\count255}
        \multiply\count255 by-60\advance\count255 by\time
   \xdef\oramin{\oramin:\ifnum\count255<10 0\fi\the\count255}}
\def\ora{\oramin }

\def\data{\number\day/\ifcase\month\or gennaio \or febbraio \or marzo \or
aprile \or maggio \or giugno \or luglio \or agosto \or settembre
\or ottobre \or novembre \or dicembre \fi/\number\year;\ \ora}

\setbox200\hbox{$\scriptscriptstyle \data $}


\newcount\pgn \pgn=1
\newcount\firstpage

\def\foglio{\number\numsec:\number\pgn\global\advance\pgn by 1}
\def\foglioa{A\number\numsec:\number\pgn\global\advance\pgn by 1}

\def\pagina{\vfill\eject}
\def\ppagina{\ifodd\pageno\pagina\null\pagina\else\pagina\fi}
\def\ppaginan{\ifodd-\pageno\pagina\null\pagina\else\pagina\fi}

\def\setind{\firstpage=\pageno}
\def\setcap#1{\null\def\titlecap{#1}\global\firstpage=\pageno}
\def\titletesi{Indici critici per sistemi fermionici in una dimensione}

\ifnum\stile=1
  \def\pagenumbers{\headline={%
  \ifnum\pageno=\firstpage\hfil\else%
     \ifodd\pageno\hfill{\sc\titlecap}~~{\bf\folio}%
      \else{\bf\folio}~~{\sc\titletesi}\hfill\fi\fi}
  \footline={\ifnum\bz=0
                   \hfill\else\rlap{\hbox{\copy200}\ $\st[\foglio]$}\hfill\fi}}
  \def\pagenumbersind{\headline={%
  \ifnum\pageno=\firstpage\hfil\else%
    \ifodd\pageno\hfill{\rm\romannumeral-\pageno}%
     \else{\rm\romannumeral-\pageno}\hfill\fi\fi}
  \footline={\ifnum\bz=0
                   \hfill\else\rlap{\hbox{\copy200}\ $\st[\foglio]$}\hfill\fi}}
\else
  \def\pagenumbers{\headline={\hfill}
     \footline={\ifnum\bz=0\hfill\folio\hfill
                \else\rlap{\hbox{\copy200}\ $\st[\foglio]$}
		   \hfill\folio\hfill\fi}}
\fi

\pagenumbers

\def\numeropag#1{
   \ifnum #1<0 \romannumeral -#1\else \number #1\fi
   }


\global\newcount\numsec\global\newcount\numfor
\global\newcount\numfig\global\newcount\numpar
\global\newcount\numteo\global\newcount\numlem

\numfig=1\numsec=0

\gdef\profonditastruttura{\dp\strutbox}
\def\senondefinito#1{\expandafter\ifx\csname#1\endcsname\relax}
\def\SIA #1,#2,#3 {\senondefinito{#1#2}%
\expandafter\xdef\csname#1#2\endcsname{#3}\else%
\write16{???? ma #1,#2 e' gia' stato definito !!!!}\fi}
\def\etichetta(#1){(\veroparagrafo.\veraformula)
\SIA e,#1,(\veroparagrafo.\veraformula)
 \global\advance\numfor by 1
\write15{\string\FU (#1){\equ(#1)}}
\9{ \write16{ EQ \equ(#1) == #1  }}}
\def \FU(#1)#2{\SIA fu,#1,#2 }
\def\etichettaa(#1){(A\veroparagrafo.\veraformula)
 \SIA e,#1,(A\veroparagrafo.\veraformula)
 \global\advance\numfor by 1
\write15{\string\FU (#1){\equ(#1)}}
\9{ \write16{ EQ \equ(#1) == #1  }}}
\def \FU(#1)#2{\SIA fu,#1,#2 }
\def\tetichetta(#1){\veroparagrafo.\veroteorema
\SIA e,#1,{\veroparagrafo.\veroteorema}
\global\advance\numteo by1
\write15{\string\FU (#1){\equ(#1)}}%
\9{\write16{ EQ \equ(#1) == #1}}}
\def\tetichettaa(#1){A\veroparagrafo.\veroteorema
\SIA e,#1,{A\veroparagrafo.\veroteorema}
\global\advance\numteo by1
\write15{\string\FU (#1){\equ(#1)}}%
\9{\write16{ EQ \equ(#1) == #1}}}
\def\letichetta(#1){\veroparagrafo.\verolemma
\SIA e,#1,{\veroparagrafo.\verolemma}
\global\advance\numlem by1
\write15{\string\FU (#1){\equ(#1)}}%
\9{\write16{ EQ \equ(#1) == #1}}}
\def\getichetta(#1){
 \SIA e,#1,{\verafigura}
 \global\advance\numfig by 1
\write15{\string\FU (#1){\equ(#1)}}
\9{ \write16{ Fig. \equ(#1) ha simbolo  #1  }}}

\def\veroparagrafo{\number\numsec}\def\veraformula{\number\numfor}
\def\verafigura{\number\numfig}\def\veroteorema{\number\numteo}
\def\verolemma{\number\numlem}

\def\geq(#1){\getichetta(#1)\galato(#1)}
\def\Eq(#1){\eqno{\etichetta(#1)\alato(#1)}}
\def\eq(#1){&\etichetta(#1)\alato(#1)}
\def\Eqa(#1){\eqno{\etichettaa(#1)\alato(#1)}}
\def\eqa(#1){&\etichettaa(#1)\alato(#1)}
\def\teq(#1){\tetichetta(#1)\talato(#1)}
\def\teqa(#1){\tetichettaa(#1)\talato(#1)}
\def\leq(#1){\letichetta(#1)\talato(#1)}

\def\Eqr{\eqno(\veroparagrafo.\veraformula)\advance\numfor by 1}
\def\eqr{&(\veroparagrafo.\veraformula)\advance\numfor by 1}
\def\Eqar{\eqno(A\veroparagrafo.\veraformula)\advance\numfor by 1}
\def\eqar{&(A\veroparagrafo.\veraformula)\advance\numfor by 1}

\def\eqv(#1){\senondefinito{fu#1}$\clubsuit$#1\write16{Manca #1 !}%
\else\csname fu#1\endcsname\fi}
\def\equ(#1){\senondefinito{e#1}\eqv(#1)\else\csname e#1\endcsname\fi}


\newdimen\gwidth

\ifnum\bz=1

\def\commenta#1{\strut \vadjust{\kern-\profonditastruttura
 \vtop to \profonditastruttura{\baselineskip
 \profonditastruttura\vss
 \rlap{\kern\hsize\kern0.1truecm
  \vbox{\hsize=1.7truecm\raggedright\nota\noindent #1}}}}}
\def\talato(#1){\strut \vadjust{\kern-\profonditastruttura
 \vtop to \profonditastruttura{\baselineskip
 \profonditastruttura\vss
 \rlap{\kern-1.2truecm{$\scriptstyle#1$}}}}}
\def\alato(#1){
 {\vtop to \profonditastruttura{\baselineskip
 \profonditastruttura\vss
 \rlap{\kern-\hsize\kern-1.2truecm{$\scriptstyle#1$}}}}}
\def\galato(#1){ \gwidth=\hsize 
 {\vtop to \profonditastruttura{\baselineskip
 \profonditastruttura\vss
 \rlap{\kern-\gwidth\kern-1.2truecm{$\scriptstyle#1$}}}}}

\else

\def\alato(#1){}
\def\galato(#1){}
\def\talato(#1){}
\def\commenta#1{}

\fi


\newskip\ttglue

\font\ftitolo=cmbx12 
\font\eighttt=cmtt8 \font\sevenit=cmti7  \font\sevensl=cmsl8
\font\sc=cmcsc10


\def\settepunti{\def\rm{\fam0\sevenrm}
\textfont0=\sevenrm \scriptfont0=\fiverm \scriptscriptfont0=\fiverm
\textfont1=\seveni \scriptfont1=\fivei   \scriptscriptfont1=\fivei
\textfont2=\sevensy \scriptfont2=\fivesy   \scriptscriptfont2=\fivesy
\textfont3=\tenex \scriptfont3=\tenex   \scriptscriptfont3=\tenex
\textfont\itfam=\sevenit  \def\it{\fam\itfam\sevenit}%
\textfont\slfam=\sevensl  \def\sl{\fam\slfam\sevensl}%
\textfont\ttfam=\eighttt  \def\tt{\fam\ttfam\eighttt}
\textfont\bffam=\sevenbf  \scriptfont\bffam=\fivebf
\scriptscriptfont\bffam=\fivebf  \def\bf{\fam\bffam\sevenbf}%
\tt \ttglue=.5em plus.25em minus.15em
\setbox\strutbox=\hbox{\vrule height6.5pt depth1.5pt width0pt}%
\normalbaselineskip=8pt\let\sc=\fiverm \normalbaselines\rm}

\let\nota=\settepunti


\font\tenmib=cmmib10
\font\sevenmib=cmmib10 scaled 800

\textfont5=\tenmib  \scriptfont5=\sevenmib  \scriptscriptfont5=\fivei

\mathchardef\aaa= "050B
\mathchardef\xxx= "0518
\mathchardef\oo = "0521
\mathchardef\Dp = "0540
\mathchardef\H  = "0548
\mathchardef\FFF= "0546
\mathchardef\ppp= "0570
\mathchardef\nnn= "0517

\newdimen\xshift \newdimen\xwidth \newdimen\yshift \newdimen\ywidth
\newdimen\laln

\def\ins#1#2#3{\nointerlineskip\vbox to0pt {\kern-#2 \hbox{\kern#1 #3}
\vss}}

\def\eqfig#1#2#3#4#5{
\hfill\break\penalty 10000\xwidth=#1 \xshift=\hsize \advance\xshift 
by-\xwidth \divide\xshift by 2
\yshift=#2 \divide\yshift by 2
\line{\hglue\xshift \vbox to #2{\vfil 
#3 \includegraphics{#4.ps}
}\hfill\raise\yshift\hbox{#5}}}

\def\eqfigbis#1#2#3#4#5#6#7{
\par\xwidth=#1 \multiply\xwidth by 2 
\xshift=\hsize \advance\xshift 
by-\xwidth \divide\xshift by 3
\yshift=#2 \divide\yshift by 2
\ywidth=#2
\line{\hfill
\vbox to \ywidth{\vfil #3 \includegraphics{#4.ps}}
\hglue20pt
\vbox to \ywidth{\vfil \includegraphics{#6.ps} #5}
\hfill\raise\yshift\hbox{#7}}}

\def\dimenfor#1#2{\par\xwidth=#1 \multiply\xwidth by 2 
\xshift=\hsize \advance\xshift 
by-\xwidth \divide\xshift by 3
\divide\xwidth by 2 
\yshift=#2 
\ywidth=#2}


\def\eqfigter#1#2#3#4#5#6#7{
\line{\hglue\xshift 
\vbox to \ywidth{\vfil #1 \includegraphics{#2.ps}}
\hglue30pt
\vbox to \ywidth{\vfil #3 \includegraphics{#4.ps}}\hfill}
\multiply\xshift by 3 \advance\xshift by \xwidth \divide\xshift by 2
\line{\hfill\hbox{#7}}
\line{\hglue\xshift 
\vbox to \ywidth{\vfil #5 \includegraphics{#6.ps}}}}


\def\8{\write13}


\def\gnuin #1 #2 #3 #4 #5 #6{\midinsert\vbox{\vbox to 260pt{
\hbox to 420pt{
\hbox to 200pt{\hfill\nota (a)\hfill}\hfill
\hbox to 200pt{\hfill\nota (b)\hfill}}
\vbox to 110pt{\vfill\hbox to 420pt{
\hbox to 200pt{\includegraphics{#1.ps}\hfill}\hfill
\hbox to 200pt{\includegraphics{#2.ps}\hfill}
}}\vfill
\hbox to 420pt{
\hbox to 200pt{\hfill\nota (c)\hfill}\hfill
\hbox to 200pt{\hfill\nota (d)\hfill}}
\vbox to 110pt{\vfill\hbox to 420pt{
\hbox to 200pt{\includegraphics{#3.ps}\hfill}\hfill
\hbox to 200pt{\includegraphics{#4.ps}\hfill}
}}\vfill\geq(#5)}
\vskip0.25cm
\0\didascalia{Fig. \equ(#5): #6}}
\endinsert}

\def\gnuinf #1 #2 #3 #4 #5 #6{\midinsert\nointerlineskip\vbox to 260pt{
\hbox to 420pt{
\hbox to 200pt{\hfill\nota (a)\hfill}\hfill
\hbox to 200pt{\hfill\nota (b)\hfill}}
\vbox to 110pt{\vfill\hbox to 420pt{
\hbox to 200pt{\includegraphics{#1.ps}\hfill}\hfill
\hbox to 200pt{\includegraphics{#2.ps}\hfill}
}}\vfill
\hbox to 420pt{
\hbox to 200pt{\hfill\nota (c)\hfill}\hfill
\hbox to 200pt{\hfill\nota (d)\hfill}}
\vbox to 110pt{\vfill\hbox to 420pt{
\hbox to 200pt{\includegraphics{#3.ps}\hfill}\hfill
\hbox to 200pt{\includegraphics{#4.ps}\hfill}
}}\vfill\geq(#5)}
\?
\0\didascalia{Fig. \equ(#5): #6}
\endinsert
\global\setbox150\vbox{\unvbox150 \*\*\0 Fig. \equ(#5): #6}
\global\setbox149\vbox{\unvbox149 \*\*
    \vbox{\centerline{Fig. \equ(#5)(a)} \nobreak
    \vbox to 200pt{\vfill\includegraphics{#1.ps_f}}}\*\*
    \vbox{\centerline{Fig. \equ(#5)(b)}\nobreak
    \vbox to 200pt{\vfill\includegraphics{#2.ps_f}}}\*\*
    \vbox{\centerline{Fig. \equ(#5)(c)}\nobreak
    \vbox to 200pt{\vfill\includegraphics{#3.ps_f}}}\*\*
    \vbox{\centerline{Fig. \equ(#5)(d)}\nobreak
    \vbox to 200pt{\vfill\includegraphics{#4.ps_f}}}
}}

\def\gnuins #1 #2 #3{\midinsert\nointerlineskip
\vbox{\line{\vbox to 220pt{\vfill
\includegraphics{#1.ps}}\hfill\geq(#2)}
\*
\0\didascalia{Fig. \equ(#2): #3}}
\endinsert}

\def\gnuinsf #1 #2 #3{\midinsert\nointerlineskip
\vbox{\line{\vbox to 220pt{\vfill
\includegraphics{#1.ps}}\hfill\geq(#2)}
\*
\0\didascalia{Fig. \equ(#2): #3}}\endinsert
\global\setbox150\vbox{\unvbox150 \*\*\0 Fig. \equ(#2): #3}
\global\setbox149\vbox{\unvbox149 \*\* 
    \vbox{\centerline{Fig. \equ(#2)}\nobreak
    \vbox to 220pt{\vfill\includegraphics{#1.ps_f}}}} 
}


\def\9#1{\ifnum\aux=1#1\else\relax\fi}
\let\numero=\number
\def\boh{\hbox{$\clubsuit$}\write16{Qualcosa di indefinito a pag. \the\pageno}}
\def\didascalia#1{\vbox{\nota\0#1\hfill}\vskip0.3truecm}
\def\frac#1#2{{#1\over #2}}
\def\V#1{{\underline{#1}}}
	\let\ciao=\bye
		\let\i=\infty
		
 	\let\0=\noindent
\def\guida{\leaders\hbox to 1em{\hss.\hss}\hfill}
\def\tende#1{\vtop{\ialign{##\crcr\rightarrowfill\crcr
              \noalign{\kern-1pt\nointerlineskip}
              \hglue3.pt${\scriptstyle #1}$\hglue3.pt\crcr}}}
\def\otto{{\kern-1.truept\leftarrow\kern-5.truept\to\kern-1.truept}}

\def\={{ \; \equiv \; }}		
\ifnum\mgnf=0
    \def\openone{\leavevmode\hbox{\ninerm 1\kern-3.3pt\tenrm1}}%
\fi
\ifnum\mgnf=1
     \def\openone{\leavevmode\hbox{\ninerm 1\kern-3.6pt\tenrm1}}%
\fi

\def\2{{1\over2}}

\def\igb{
    \mathop{\raise4.pt\hbox{\vrule height0.2pt depth0.2pt width6.pt}
    \kern0.3pt\kern-9pt\int}}

\def\st{\scriptscriptstyle}
\let\\=\noindent
\def\*{\vskip0.5truecm}
\def\?{\vskip0.75truecm}
\def\item#1{\vskip0.1truecm\parindent=0pt\par\setbox0=\hbox{#1\ }
     \hangindent\wd0\hangafter 1 #1 \parindent=\stdindent}
\def\sit{\vskip0.1truecm}


\def\ie{\hbox{\sl i.e.\ }}

\def\qed{\hfill\break\nobreak\vbox{\vglue.25truecm\line{\hfill\raise1pt 
          \hbox{\vrule height9pt width5pt depth0pt}}}\vglue.25truecm}


\def\gint(#1)(#2)(#3){{\cal D}#1^{#2}\,e^{(#1^{#2+},#3#1^{#2-})}}

  \def\V0{{\bf 0}}    \def\tt{{\bf t}}


\ifnum\cind=1
\def\prtindex#1{\immediate\write\indiceout{\string\parte{#1}{\the\pageno}}}
\def\capindex#1#2{\immediate\write\indiceout{\string\capitolo{#1}{#2}{\the\pageno}}}
\def\parindex#1#2{\immediate\write\indiceout{\string\paragrafo{#1}{#2}{\the\pageno}}}
\def\subindex#1#2{\immediate\write\indiceout{\string\sparagrafo{#1}{#2}{\the\pageno}}}
\def\appindex#1#2{\immediate\write\indiceout{\string\appendice{#1}{#2}{\the\pageno}}}
\def\paraindex#1#2{\immediate\write\indiceout{\string\paragrafoapp{#1}{#2}{\the\pageno}}}
\def\subaindex#1#2{\immediate\write\indiceout{\string\sparagrafoapp{#1}{#2}{\the\pageno}}}
\def\bibindex#1{\immediate\write\indiceout{\string\bibliografia{#1}{Bibliografia}{\the\pageno}}}
\def\preindex#1{\immediate\write\indiceout{\string\premessa{#1}{\the\pageno}}}
\else
\def\prtindex#1{}
\def\capindex#1#2{}
\def\parindex#1#2{}
\def\subindex#1#2{}
\def\appindex#1#2{}
\def\paraindex#1#2{}
\def\subaindex#1#2{}
\def\bibindex#1{}
\def\preindex#1{}
\fi

\def\leaderfill{\leaders\hbox to 1em{\hss . \hss} \hfill }


\newdimen\capsalto \capsalto=0pt
\newdimen\parsalto \parsalto=20pt
\newdimen\sparsalto \sparsalto=30pt
\newdimen\tratitoloepagina \tratitoloepagina=2\parsalto
\def\aboveparteskip{\bigskip \bigskip}
\def\belowparteskip{\medskip \medskip}
\def\abovecapitskip{\bigskip}
\def\belowcapitskip{\medskip}
\def\belowparskip{\smallskip}
%


\def\parte#1#2{
   \9{\immediate\write16
      {#1     pag.\numeropag{#2} }}
   \aboveparteskip 
   \noindent 
   {\ftitolo #1} 
   \hfill {\ftitolo \numeropag{#2}}\par
   \belowparteskip
   }


\def\premessa#1#2{
   \9{\immediate\write16
      {#1     pag.\numeropag{#2} }}
   \abovecapitskip 
   \noindent 
   {\it #1} 
   \hfill {\rm \numeropag{#2}}\par
   \belowcapitskip
   }


\def\bibliografia#1#2#3{
  \ifnum\stile=1
   \9{\immediate\write16
      {Bibliografia    pag.\numeropag{#3} }}
   \belowcapitskip
   \noindent 
   {\bf Bibliografia} 
   \hfill {\bf \numeropag{#3}}\par
  \else
    \paragrafo{#1}{References}{#3}
\fi
   }


\newdimen\newstrutboxheight
\newstrutboxheight=\baselineskip
\advance\newstrutboxheight by -\dp\strutbox
\newdimen\newstrutboxdepth
\newstrutboxdepth=\dp\strutbox
\newbox\newstrutbox
\setbox\newstrutbox = \hbox{\vrule 
   height \newstrutboxheight 
   width 0pt 
   depth \newstrutboxdepth 
   }
\def\newstrut
   {\relax \ifmmode \copy \newstrutbox \else \unhcopy \newstrutbox \fi}
%
%
\vfuzz=3.5pt
%
%
\newdimen\indexsize \indexsize=\hsize
\advance \indexsize by -\tratitoloepagina
\newdimen\dummy
\newbox\parnum
\newbox\parbody
\newbox\parpage
%

%

\def\mastercap#1#2#3#4#5{
   \9{\immediate\write16
      {Cap. #3:#4     pag.\numeropag{#5} }}
   \abovecapitskip
   \setbox\parnum=\hbox {\kern#1\newstrut{#2}
			{\bf Capitolo~\number#3.}~}
   \dummy=\indexsize
   \advance\indexsize by -\wd\parnum
   \setbox\parbody=\vbox {
      \hsize = \indexsize \noindent \newstrut 
      {\bf #4}
      \newstrut \hss}
   \indexsize=\dummy
   \setbox\parnum=\vbox to \ht\parbody {
      \box\parnum
      \vfill 
      }
   \setbox\parpage = \hbox to \tratitoloepagina {
      \hss {\bf \numeropag{#5}}}
   \noindent \box\parnum\box\parbody\box\parpage\par
   \belowcapitskip
   }
\def\capitolo#1#2#3{\mastercap{\capsalto}{}{#1}{#2}{#3}}
%


%
\def\masterapp#1#2#3#4#5{
   \9{\immediate\write16
      {App. #3:#4     pag.\numeropag{#5} }}
   \abovecapitskip
   \setbox\parnum=\hbox {\kern#1\newstrut{#2}
			{\bf Appendice~A\number#3:}~}
   \dummy=\indexsize
   \advance\indexsize by -\wd\parnum
   \setbox\parbody=\vbox {
      \hsize = \indexsize \noindent \newstrut 
      {\bf #4}
      \newstrut \hss}
   \indexsize=\dummy
   \setbox\parnum=\vbox to \ht\parbody {
      \box\parnum
      \vfill 
      }
   \setbox\parpage = \hbox to \tratitoloepagina {
      \hss {\bf \numeropag{#5}}}
   \noindent \box\parnum\box\parbody\box\parpage\par
   \belowcapitskip
   }
\def\appendice#1#2#3{\masterapp{\capsalto}{}{#1}{#2}{#3}}
%

%

\def\masterpar#1#2#3#4#5{
   \9{\immediate\write16
      {par. #3:#4     pag.\numeropag{#5} }}
   \setbox\parnum=\hbox {\kern#1\newstrut{#2}\number#3.~}
   \dummy=\indexsize
   \advance\indexsize by -\wd\parnum
   \setbox\parbody=\vbox {
      \hsize = \indexsize \noindent \newstrut 
      #4
      \newstrut \hss}
   \indexsize=\dummy
   \setbox\parnum=\vbox to \ht\parbody {
      \box\parnum
      \vfill 
      }
   \setbox\parpage = \hbox to \tratitoloepagina {
      \hss \numeropag{#5}}
   \noindent \box\parnum\box\parbody\box\parpage\par
   \belowparskip
   }
\def\paragrafo#1#2#3{\masterpar{\parsalto}{}{#1}{#2}{#3}}
\def\sparagrafo#1#2#3{\masterpar{\sparsalto}{}{#1}{#2}{#3}}
%


%
\def\masterpara#1#2#3#4#5{
   \9{\immediate\write16
      {par. #3:#4     pag.\numeropag{#5} }}
   \setbox\parnum=\hbox {\kern#1\newstrut{#2}A\number#3.~}
   \dummy=\indexsize
   \advance\indexsize by -\wd\parnum
   \setbox\parbody=\vbox {
      \hsize = \indexsize \noindent \newstrut 
      #4
      \newstrut \hss}
   \indexsize=\dummy
   \setbox\parnum=\vbox to \ht\parbody {
      \box\parnum
      \vfill 
      }
   \setbox\parpage = \hbox to \tratitoloepagina {
      \hss \numeropag{#5}}
   \noindent \box\parnum\box\parbody\box\parpage\par
   \belowparskip
   }
\def\paragrafoapp#1#2#3{\masterpara{\parsalto}{}{#1}{#2}{#3}}
\def\sparagrafoapp#1#2#3{\masterpara{\sparsalto}{}{#1}{#2}{#3}}
%


\ifnum\stile=1

\def\newcap#1{\setcap{#1}
\vskip2.truecm\advance\numsec by 1
\\{\ftitolo \numero\numsec. #1}
\capindex{\numero\numsec}{#1}
\vskip1.truecm\numfor=1\pgn=1\numpar=1\numteo=1\numlem=1
}

\def\newapp#1{\setcap{#1}
\vskip2.truecm\advance\numsec by 1
\\{\ftitolo A\numero\numsec. #1}
\appindex{A\numero\numsec}{#1}
\vskip1.truecm\numfor=1\pgn=1\numpar=1\numteo=1\numlem=1
}

\def\newpar#1{
\vskip1.truecm
\vbox{
\\{\bf \numero\numsec.\numero\numpar. #1}
\parindex{\numero\numsec.\numero\numpar}{#1}
\*{}}
\nobreak
\advance\numpar by 1
}

\def\newpara#1{
\vskip1.truecm
\vbox{
\\{\bf A\numero\numsec.\numero\numpar. #1}
\paraindex{\numero\numsec.\numero\numpar}{#1}
\*{}}
\nobreak
\advance\numpar by 1
}

\else

\def\newsec#1{\vskip1.truecm
\advance\numsec by 1
\vbox{
\\{\bf \numero\numsec. #1}
\parindex{\numero\numsec}{#1}
\*{}}\numfor=1\pgn=1\numpar=1\numteo=1\numlem=1
\nobreak
}

\def\newapp#1{\vskip1.truecm
\advance\numsec by 1
\vbox{
\\{\bf A\numero\numsec. #1}
\appindex{A\numero\numsec}{#1}
\*{}}\numfor=1\pgn=1\numpar=1\numteo=1\numlem=1
}

\def\appendices{\numsec=0\def\Eq(##1){\Eqa(##1)}\def\eq(##1){\eqa(##1)}
\def\teq(##1){\teqa(##1)}}

\def\biblio{\vskip1.truecm
\vbox{
\\{\bf References.}\*{}
\bibindex{{}}}\nobreak\makebiblio
}

\fi


\newread\indicein
\newwrite\indiceout

\def\faindice{
\openin\indicein=\jobname.ind
\ifeof\indicein\relax\else{
\ifnum\stile=1
  \pagenumbersind
  \pageno=-1
  \setind
  \null
  \vskip 2.truecm
  \\{\ftitolo Indice}
  \vskip 1.truecm
  \parskip = 0pt
  \input \jobname.ind
  \ppaginan
\else
\\{\bf Index}
\*{}
 \input \jobname.ind
\fi}\fi
\closein\indicein
\def\nomeindice{\jobname.ind}
\immediate\openout \indiceout = \nomeindice
}


\newwrite\bib
\immediate\openout\bib=\jobname.bib
\global\newcount\bibex
\bibex=0
\def\verabib{\number\bibex}

\ifnum\tipobib=0
\def\cita#1{\expandafter\ifx\csname c#1\endcsname\relax
\hbox{$\clubsuit$}#1\write16{Manca #1 !}%
\else\csname c#1\endcsname\fi}
\def\rife#1#2#3{\immediate\write\bib{\string\raf{#2}{#3}{#1}}
\immediate\write15{\string\C(#1){[#2]}}
\setbox199=\hbox{#2}\ifnum\maxit < \wd199 \maxit=\wd199\fi}
\else
\def\cita#1{%
\expandafter\ifx\csname d#1\endcsname\relax%
\expandafter\ifx\csname c#1\endcsname\relax%
\hbox{$\clubsuit$}#1\write16{Manca #1 !}%
\else\probib(ref. numero )(#1)%
\csname c#1\endcsname%
\fi\else\csname d#1\endcsname\fi}%
\def\rife#1#2#3{\immediate\write15{\string\Cp(#1){%
\string\immediate\string\write\string\bib{\string\string\string\raf%
{\string\verabib}{#3}{#1}}%
\string\Cn(#1){[\string\verabib]}%
\string\CCc(#1)%
}}}%
\fi

\def\Cn(#1)#2{\expandafter\xdef\csname d#1\endcsname{#2}}
\def\CCc(#1){\csname d#1\endcsname}
\def\probib(#1)(#2){\global\advance\bibex+1%
\9{\immediate\write16{#1\verabib => #2}}%
}

\def\C(#1)#2{\SIA c,#1,{#2}}
\def\Cp(#1)#2{\SIAnx c,#1,{#2}}

\def\SIAnx #1,#2,#3 {\senondefinito{#1#2}%
\expandafter\def\csname#1#2\endcsname{#3}\else%
\write16{???? ma #1,#2 e' gia' stato definito !!!!}\fi}

\bibskip=10truept
\def\hboxto{\hbox to}

\catcode`\{=12\catcode`\}=12
\catcode`\<=1\catcode`\>=2
\immediate\write\bib<
	\string\halign{\string\hboxto \string\maxit%
	{\string #\string\hfill}&%
        \string\vtop{\string\parindent=0pt\string\advance\string\hsize%
	by -1.55truecm%
	\string#\string\vskip \bibskip
	}\string\cr%
>
\catcode`\{=1\catcode`\}=2
\catcode`\<=12\catcode`\>=12

\def\raf#1#2#3{\ifnum \bz=0 [#1]&#2 \cr\else
\llap{${}_{\rm #3}$}[#1]&#2\cr\fi}

\newread\bibin

\catcode`\{=12\catcode`\}=12
\catcode`\<=1\catcode`\>=2
\def\chiudibib<
\catcode`\{=12\catcode`\}=12
\catcode`\<=1\catcode`\>=2
\immediate\write\bib<}>
\catcode`\{=1\catcode`\}=2
\catcode`\<=12\catcode`\>=12
>
\catcode`\{=1\catcode`\}=2
\catcode`\<=12\catcode`\>=12

\def\makebiblio{
\ifnum\tipobib=0
\advance \maxit by 10pt
\else
\maxit=1.truecm
\fi
\chiudibib
\immediate \closeout\bib
\openin\bibin=\jobname.bib
\ifeof\bibin\relax\else
\raggedbottom
\input \jobname.bib
\fi
}

\openin13=#1.aux \ifeof13 \relax \else
\input #1.aux \closein13\fi
\openin14=\jobname.aux \ifeof14 \relax \else
\input \jobname.aux \closein14 \fi
\immediate\openout15=\jobname.aux

\def\<{\langle}
\def\>{\rangle}

\def\V#1{{\underline{#1}}}


\titolo{On the Validity of the Conjugate Pairing Rule for Lyapunov Exponents}
\autore{F. Bonetto}{Rockefeller University. Present address: Mathematics 
Department, Rutgers University} 
\autore{E.G.D. Cohen}{Rockefeller University} 
\autore{C. Pugh}{Mathematics Department University of California 
Berkeley California, 94720}

\abstract{For Hamiltonian systems subject to an external
potential, which in the presence of a thermostat will reach a
nonequilibrium stationary state, Dettmann and Morriss$^{\cita{DM}}$
proved a strong conjugate pairing rule (SCPR) for pairs of Lyapunov
exponents in the case of isokinetic (IK) stationary states which have
a given kinetic energy.  This SCPR holds for all initial phases of the
system, all times t and all numbers of particles N. This proof was
generalized by Wojtkovski and Liverani$^{\cita{LiWo}}$ to include hard
interparticle potentials. A geometrical reformulation of those results
is presented. The present paper proves numerically, using periodic
orbits for the Lorentz gas, that SCPR cannot hold for isoenergetic
(IE) stationary states, which have a given total internal energy.  In
that case strong evidence is obtained for CPR to hold for large N and
t, where it can be conjectured that the larger N, the smaller t will
be. This suffices for statistical mechanics.}

\parole{dynamical systems, Lyapunov exponents, symmetry, pairing rule}

\prima

\newsec{Introduction.} 

This paper consists of three parts: a statistical mechanical, a
numerical and a mathematical part. They are not equally accessible to
readers interested in the new and developing interaction between
dynamical systems theory and statistical mechanics of dissipative
systems. This reflects, in our opinion, the multidisciplinary approach
involving on the one hand the mathematical theory of dynamical systems
and on the other hand the physical theory of dissipative macroscopic,
\ie , statistical mechanical systems. In order to make progress towards
a deeper understanding of the dynamical foundation of the statistical
mechanics of dissipative macroscopic systems, such a combined approach
seems unavoidable. However, it leads, regrettably, to difficulties in
understanding the language of the different disciplines.

Since this paper is primarily adressed to physicists, we have written
the core of the paper in a language that should be accessible to a
large portion of them. However, an interesting mathematical
unification and reformulation of previous results has been obtained by
one of us (C.P.). This work is of a more mathematical nature than the
rest of the paper, but make a connection between mathematics and
physics, in particular the mathematics of symplectic spaces and the
physics of thermostatted dynamical systems. For that reason we have
summarized the main results of this work in a Discussion Remark, but
reserved a sketch of the details to an Appendix.

The modelling of dissipative macroscopic systems by dynamical systems,
i.e., by sets of equations of motion for the particles of the system,
which mimic the behavior of such systems under the influence of an
external field in a nonequilibrium stationary state, can be achieved
by adding a suitable dynamical thermostatting term to the equations of
motion. In fact, in all computer simulations the dynamical system
reaches a stationary state, by adjusting a thermostat term in the
equations of motion in such a way that the heat developed due to the
work done on the system by the external forces is removed via effective
friction terms in the equations of motion of the system.  This
approach has raised a lot of interest in the last ten years.  In
particular in 1988, Posch and Hoover$^{\cita{PH}}$ found a
relationship between the average phase space contraction rate of a
many particle color diffusion system under the influence of an
external color field and the work done on the particles of the system,
if the internal energy of the system in the stationary state was kept
constant (iso-energetic (IE) thermostat).  Another way of saying this
is that in the stationary state the average phase
space contraction rate of the system is equal to the irreversible
entropy production rate in the system$^{\cita{CR}}$.  This allowed one
to express transport coefficients in terms of the sum of the Lyapunov
exponents of the system and established a direct relation between
dynamical systems theory in phase space and macroscopic physics in
ordinary space.  Similar relations have been found for other dynamical
systems, including those representing viscous flow, if the stationary
state is one with a fixed kinetic or internal energy.  In the case of
hard core interparticle interactions, when no interparticle potential
energy is present, iso-energetic and iso-kinetic (IK) systems, which
have constant kinetic energy, are the same.  Most computer experiments
have been carried out for the IK case and some for the IE case as
well$^{\cita{EM},\cita{SE}}$.

The dynamical equations describing the macroscopic systems were solved
numerically, using nonequilibrium molecular dynamics (NEMD). It was
found this way that an important simplification occurred - for example
in the computation of the transport coefficients - in that the sum of
all Lyapunov exponents, which determines the phase space contraction
rate, obeys a conjugate pairing rule (CPR) (see \cita{ECM1}). 
That is, the sum of every
pair of conjugate Lyapunov exponents adds up to the same (negative)
constant, which is equal to the average phase space contraction rate
of the system in the stationary state, which in turn was equated to
the average entropy production rate of the system.  Equivalently, one
can say that the Lyapunov spectrum of the system is symmetric.

A mathematical proof of the CPR for the case of a possibly
time-dependent, i.e., non-autonomous, Hamiltonian system with a
constant friction added to the equations of motion, was given
independently of the above developments in 1988 by
Dressler$^{\cita{D}}$.  The system she considered in this case was
neither IK or IE.

For an IK Hamiltonian system consisting of N particles $i =1,...,N$
with a Gaussian thermostat, which keeps the kinetic energy
constant$^{\cita{EM}}$, one has in three dimensions $(d = 3)$ the
Hamiltonian:
$$H(\V p, \V q) = \sum^N_{i=1} \frac{\vec{p}^2_i}{2m} + \phi^{int} (\V q)
+ \phi^{ext}(\V q)\Eq(Hami)
$$
with the equations of motion
$$\left\{\eqalign{
\dot{\vec{q}}_i &= \frac{\vec{p}_i}{m}\cr
\dot{\vec{p}}_i &= \vec f_i^{int}(\V q) + \vec f_i^{ext}(\V q)) - 
\alpha(\V p, \V q) {\vec{p}}_i\cr}\right.\Eq(moti)
$$
where $\alpha(\V p, \V q)=\alpha_{IK}(\V p, \V q)$ is given by
$$
\alpha_{IK}(\V p, \V q) =  \frac{\sum_i \biggl({\vec{p}}_i , 
\bigl[\vec f_i^{int}(\V q) + \vec f_i^{ext}(\V q))\bigr]\biggr)}
{\sum_i {\vec{p}}_i^{2}}\Eq(aIK)
$$
where $\phi^{int}(\V q)$ and $\phi^{ext}(\V q)$ are the potential
energy of the particles of the system and with the external field,
respectively, where $\V p = ({\vec{p}}_1, ..., {\vec{p}}_N$) and $\V q
= ({\vec{q}}_1, ...., {\vec{q}}_N)$ , ${\vec{p}}_i=(p_x,p_y,p_z)$ and
${\vec{q}}_i=(q_x,q_y,q_z)$ are three dimensional vectors, $\vec
f^{int}_i(\V q) = \partial \phi^{int}(\V q)/\partial \vec{q}_i$, $\vec
f^{ext}_i(\V q) = \partial \phi^{ext}(\V q)/\partial \vec{q}_i$, and
$(\cdot,\cdot)$ represent the usual scalar product in three
dimensions. $\alpha_{IK}(\V p, \V q)$ is the thermostatting function
(or effective friction) which will be determined in such a way that
the total kinetic energy of the system remains constant in time.  It
is a mechanical representation of the heat exchange via the boundaries
of the system representing the thermostat in a real system.

Dettmann and Morriss$^{\cita{DM}}$ recently proved a much stronger CPR
than observed in the stationary states of the computer simulations:
their CPR holds for every initial condition ${\bf x}_0 \equiv (\V p_0,
\V q_0)$, at every time $t$ and for any number $N$ of particles. We
refer to this Dettmann and Morriss pairing as a {\sl strong} CPR
(SCPR).  Isolating two zero Lyapunov exponents, the remaining $6N-2$
local Lyapunov exponents over time $t$ for any initial state and any
number $N$ of particles obey the CPR.  If the local Lyapunov exponents
are ordered as $\lambda_1 >
\lambda_2 > .... > \lambda_{6N-2}$, then adding up the conjugate pairs 
$\lambda_1 + \lambda_{6N-2}, \lambda_2 + \lambda_{6N-3}, ....,$ leads
for every such pair to a constant $- < \alpha >_t$, where $< >_t$
means a time time average of $\alpha(\V p, \V q)$ over time $t$:
$-\int^t_o \alpha (s) ds/t$.

It seems natural to extend the proof of Dettmann and Morriss from the IK to 
the IE case.  The only difference is that $\alpha (\V p, \V q)$ is not given
by $\alpha_{IK}(\V p, \V q)$ of eq.\equ(aIK) but by the simpler expression:
$$
\alpha_{IE}(\V p, \V q) = 
\frac{\sum_i \bigl({\vec{p}}_i, \vec f_i^{ext}(\V q)\bigr)}
{\sum_i {\vec{p}}_i^2}\Eq(aIE)
$$

We will call $H_0$ the constrained part of the Hamiltonian $H$, \ie

$$H^{IK}_0=\sum^N_{i=1} \frac{\vec{p}^2_i}{2m}\Eq(H0)$$
for the IK case and

$$H^{IE}_0=\sum^N_{i=1} \frac{\vec{p}^2_i}{2m} + \phi^{int} (\V q)\Eq(H00)$$
for the IE case.
The question then arises to what extent the Dettmann and Morriss' proof of
SCPR for the equations \equ(Hami)-\equ(aIK) is typical for general 
dissipative dynamical equations. 

The Dettmann and Morriss proof has recently been generalized by 
Wojt\-kow\-ski and Li\-ve\-ra\-ni$^{\cita{LiWo}}$ to include
hard core contributions to the interaction potential $\phi^{int}(\V q)$, 
for which $\vec{\nabla}_i \phi^{int}(q)$ is not defined.

The Dettmann and Morriss proof makes use of the observation that the
CPR as stated above only holds in a 2dN-2 dimensional (i.e., a
2-co-dimensional) subspace of the full 2dN-dimensional phase space of
the dynamical system.  Two zero Lyapunov exponents associated with the
conservation of kinetic energy and with the motion along the system's
phase space trajectory have first to be eliminated, in order to allow
the remaining Lyapunov exponents to pair to a different (non-zero)
value.  A similar procedure, replacing the conserved kinetic by the
total energy does not lead to a SCPR, like that found by Dettmann and
Morriss.  In fact, computer simulations for the Lorentz gas indicate
that, in general, no SCPR can be expected in this case.  On the
contrary, our computer simulations seem to indicate that the CPR may
only hold for large systems and/or long times. The reason is not clear
at the moment, but might be related to the special form of the
friction term in eq.\equ(moti): $-\alpha_{IK}(\V p, \V q)
\vec{p}_i$ which obtains both for Gaussian and Nos\'{e}-Hoover
thermostats$^{\cita{EM}}$.  For, the IK condition, $\alpha_{IK}(\V p,
\V q)$ contains $\phi^{int}(\V q)$ in addition to $\phi^{ext}(\V q)$
(the latter appearing exclusively in $\alpha_{IE}(\V p, \V q)$
(cf.eq.\equ(aIE)).  This could be responsible for the validity of the
local CPR in the IK case.  In fact, computer simulations show that the
CPR continues to be obeyed to machine precision before, during and
after a collision of one moving point particle interacting via a
Weeks-Chandler-Andersen (WCA) potential (cf.eq.\equ(WCA1)
\equ(WCA2)) with one stationary scatterer in the Lorentz gas IK case, 
but is violated considerably in the IE case.  Although it is unclear
at present why the $\alpha_{IK}(\V p, \V q)$ is able to assure the
CPR, the fact that it contains $\phi^{int}(\V q) = \phi^{WCA}(\V q)$,
the same potential that causes the scattering, could well be relevant
in this context.  On the other hand, it is very hard to see how
$\alpha_{IE}(\V p, \V q)$, which only contains the external potential,
could physically force a CPR during a collision governed by
$\phi^{int}(\V q)$.

Strictly speaking, when more than one moving particle is present in
the system, the proper thermostatting condition involves, in general,
the peculiar rather than the laboratory velocities, that appear in the
equations \equ(Hami)-\equ(aIE).  Especially in the case of viscous
flow it is important to identify the relevant particle velocities and
the temperature of the system with the peculiar rather than the actual
velocities of the particles, where the former refer to the velocity of
each particle relative to the average local flow velocity of the fluid
at the position of the particle (cf.eqs\equ(lortN)\equ(aIPK)). They
will be used in Section 4.

Our main results can be summarized as follows.

\item{1.}For a Lorentz gas, where a single point particle moving 
in a regular triangular lattice of fixed spherical WCA scatterers
under the influence of a (color) field subject to a Gaussian
thermostat to attain a stationary state, very carefully controlled
numerical simulations of period 3 to period 8 periodic orbits, have
excluded the validity of strong SCPR for an IE constraint, while being
consistent with an IK constraint (for all $t$ and $x_0$).
 
\item{2.}On the other hand both IE and the use of IK with the proper 
thermostatting condition with peculiar, rather than laboratory
velocities, for 9, 16, 32 particles moving through WCA scatterers,
strongly suggested the validity of the CPR for sufficiently large $N$
and possibly for each $t$, with $N$ large enough.  In fact, at present
it is unclear whether $N \rightarrow \infty$ would be sufficient for
the CPR to hold in general, i.e., for all $t$, or whether a large $t$
condition, i.e., $t \rightarrow \infty$, should be imposed in
addition.

The ``mechanization'' of the thermostats, as shown in the friction terms
$\sim \alpha {\vec{p}}_i$ in the above equations, was developed in numerical
simulations in the beginning of the 1980's$^{\cita{EM}}$.  This extended to heat 
exchange
the ``mechanization'' of thermal external forces - as occur in viscous or 
thermal phenomena - already introduced in linear response theory in the 
middle 1950's$^{\cita{Z}}$.  Together they form a basis for a 
connection between statistical mechanical and dynamical systems.  To
what extent these ``mechanized'' equations of motion adequately represent
the real physical systems and the application of dynamical systems theory
considerations to them reflect properties of real systems, remains, at
present, an open question.  

For large ``mechanized'' systems a number of properties of real
systems, such as the positivity of the entropy
production$^{\cita{Ru}}$ and the Onsager reciprocal
relations$^{\cita{GR}}$ have recently been established.  For a further
discussion of this point we refer to a forthcoming
paper$^{\cita{CR}}$.

If the generic case for the CPR to hold is large $N$ and/or large $t$, then 
it is in general a statistical mechanical property in a nonequilibrium
stationary state of a system, rather than a dynamical property, where it
would hold for all $N$ and/or all $t$.  

The outline of the paper is as follows.  

In section 2 various attempts are described to prove SCPR for the IE
one particle Lorentz model, where one particle moves through
scatterers placed on a regular triangular lattice.  In section 3, very
accurate controlled numerical results for a one particle regular IE
Lorentz model are presented, which contain apparently incontrovertable
evidence of a violation of the SCPR for certain periodic orbits (see
Table 2), contrary to the IK Lorentz model, as proved by Dettmann and
Morriss.  In section 4, numerical evidence for an aymptotic CPR for
large $N$ for IE and for a different thermostat (IPK), that makes sense
only for many particle systems and in which peculiar rather than
laboratory velocities appear, is presented.  Section 5 summarizes some
of the results and their implications obtained in the paper as well as
a reformulation of the strong pairing rule condition. Appendix A1 discusses
the difficulties one encounters, if one tries to map an IE system into a
Hamiltonian system, like is done in \cita{DM2} for IK systems. Finally
Appendix A2 gives a sketch of the proof of the reformulated pairing
rules condition in a geometrical form.

\newsec{Numerical results: the Lorentz model} 

In this and the following sections we report the numerical simulations we 
have done on the 
pairing problem. We focus our attention on the simplest system: a periodic 
Lorentz model with one or more moving particles through periodically
arranged scatterers. The results on a single moving particle are reported 
in this and the next section, those on many moving particles in sec. 4.

The Lorentz model we consider in this section then consists of one particle 
moving in a regular triangular array of scatterers. The particle interacts
with the scatterers via a Weeks-Chandler-Andersen (WCA)
potential and is subject to an external field and an IK or IE
thermostat. The equations of motion are thus given by eq.\equ(moti) where 
$N=1$ and, writing direcly $\vec q$, $\vec p$ instead of 
$\V q=({\vec q}_1)$, $\V p=({\vec p}_1)$,
we have $\phi^{ext}(\vec q)=(\vec q,\vec E)$, with $\vec E$ 
a fixed three dimensional vector chosen to be $\vec E=(1.1,0.1,0.1)$, 
$\phi^{int}(\vec q)=\phi^{WCA}({\vec{q}})$ where $\phi^{WCA}({\vec{q}})$ 
is the (finite range) WCA potential given by
$$\phi^{WCA}({\vec{q}})=\sum_{n\in Z^3}\tilde \phi ({\vec{q}}-{\vec{q}}_n)
\Eq(WCA1)$$
with
$${\tilde{\phi}}({\vec{q}})=\cases{{1\over \Vert {\vec{q}} \Vert^{12}}-
{1\over \Vert {\vec{q}} \Vert^{6}}+0.25& if\ $\Vert {\vec{q}}
\Vert^6<2$\cr 0& otherwise}\Eq(WCA2)$$
and ${\vec{q}}_n$ the centers of the scatterers on the regular triangular 
lattice.

This system can be considered as a model for electro-conductivity in
which the moving particle represents an electron and the scatterer
represents the ion of the metal. In the present context it is usually
used as a model for color conductivity, where the moving particle has
a color charge rather than an electrical charge$^{\cita{EM}}$, so that
(electrical) interactions between particles, when more than one is
present, are avoided.  We fix the units in such a way that the mass of
the particle is equal to 1, its charge is also equal to 1. The
scatterers have a radius of $2^{1/6}$, a distance of 2.5  
and the time the particle needs
to cover a distance 1 in free space, \ie when not interacting with a
scatterer, is $\sqrt2$.  The last condition just means that the
constant energy (kinetic or total) of the particle is fixed to 1.
From a mathematical point of view the scatterers are needed to
introduce a strong chaos in the system. In general no analytic results
are known for this system with respect to its ergodic or chaotic
properties. The closest system for which ergodicity and the existence
and uniqueness of a limiting distribution (SRB) for the Liouville measure
has been proved is the case in which the WCA scatterers are
substituted by hard scatterers and the space dimension is 2 (see
\cita{CELS}).

The number of degrees of freedom of the system is three. This means
that we have two zero Lyapunov exponents and only 4 non trivial
Lyapunov exponents: $\l_1>\l_2>\l_3>\l_4$, of which $\l_1$ and $\l_2$
are positive, $\l_3$ and $\l_4$ negative, if $\vec E$ is not too
large$^{\cita{DM2},\cita{CELS}}$. If we introduce the pairing error
by:
$$p_e=\l_1-\l_2-\l_3+\l_4\Eq(pe)$$
we have for a system for which CPR holds that $p_e=0$. This
implies that we can use $p_e$ as a measure of the discrepancy of the
Lyapunov exponents from pairing. 

Strictly speaking Lyapunov exponents are defined only in the limit
when the time length of a trajectory goes to infinity. Moreover they
are only defined almost everywhere with respect to some ergodic
measure$^{\cita{ER}}$.  In what follows we will use the term local
Lyapunov exponent to refer to a definition of a Lyapunov exponent for
a single trajectory for a finite time. The particular definition
needed will be clear from the context. We will drop the term ``local''
when no confusion can arise.

In our calculations for the one particle Lorentz model, we focussed 
our attention on three main points:

\item{A)} asymptotic pairing;

\item{B)} possibility of generalizing the DM scheme;

\item{C)} periodic orbits (see section 3);

\vskip0.1truecm
\*
\0{A. Asymptotic pairing.} 
\*
To compute the asymptotic Lyapunov exponents, we calculated the local
Lyapunov exponents for as long a time as possible, typically for about
50,000 time units.  With our units the time unit equalled the distance
unit.  We use the standard algorithm$^{\cita{BGGS}}$ consisting of
evolving a basic set of unit vectors in the tangent space with the
tangent flow and orthonormalizing them from time to time (see
\cita{ER} for a theoretical justification of this algorithm). It is
very hard to give a reliable error estimate on this kind of
computation. In fact there are two main sources of error:

\item{1)} the intrinsic instability of the dynamical system due to the
presence of two positive Lyapunov exponents (quite clear from the
simulation) prevents us from following a trajectory of the real system
for a long time. In fact even using a very accurate integration method
(we used an adaptive step size $4^{th}$ order Runge-Kutta (RK)
integrator with an error control of $10^{-11}$ at each time step) any
truncation error will be exponentially increased. The typical argument
to solve this problem is to consider the truncation error as a random
perturbation and use the ``shadowing lemma'' to argue that there must
be a trajectory of the real system near by the simulated
one$^{\cita{Bo}}$. This has, in our opinion, no real mathemathical 
basis. So we will proceed assuming that the
Lyapunov exponents of systems like (2.1) behave continuously with
respect to small perturbation-like truncation errors, even if we
cannot prove this. We refer to \cita{BGG} for further discussion.

\item{2)} the errors introduced by the numerical round-off in the
evaluation of the tangent flow, in orthonormalizing the evolved basis of
tangent vectors and the impossibility to do the infinite time limit 
on a computer. To deal with this we proceeded in two different ways. The first
one was the typical one (see e.g.\cita{PO}) and consisted in considering as 
error the variance of the computed exponent on the last part of the
trajectory. The other way was to run more than one
simulation, starting from different randomly chosen initial points.
We got essentially the same results from both these methods.
\sit

To have a clearer idea of the asymptotic values of the Lyapunov
exponents, we monitor the 6 local Lyapunov exponents as a function of time
for a single trajectory. Using the algorithm of \cita{BGGS}, we then have to 
follow these  
6 local Lyapunov exponents. In all cases, we observe that after a short
transient time two of them decrease to 0 quite regularly; these two are
related to the conservation of energy and the direction of motion. It is
natural to consider the remaining four Lyapunov exponents as the non trivial 
and possibly pairing ones and use the first two 
(which we will call the zero local Lyapunov exponents) as a check
on the precision reached in approaching the infinite time limit. 
In fact, only when these two zero Lyapunov exponents, $\l_0$ and $\l_{00}$,
have become approximately equal to zero, can CPR-like behavior be expected 
to hold for the other four asymptotic-like Lyapunov exponents. We ran
simulations both with IK and IE thermostats. Clearly we expect the
pairing error for the system with the IK thermostat to go to zero and this 
can be used as a further check of the precision of the algorithm. 
The results are reported in Fig. \equ(fig1).

We see that the three lines in Fig. \equ(fig1)(a) 
for the IK system approach zero
regularly. They can easily be described by a $C/t$ behavior with $C=1.5$ 
for the pairing error, $C=6.0$ for the largest zero exponent $\l_0$ and 
$C=0$ for the smallest exponent $\l_{00}$. Moreover, pairing is very well
satisfied, the pairing error being always smaller than the largest zero
exponent $\l_0$. All this is clearly in good agreement with
\cita{DM} and reflects the fact that, if one
had used a proper Poincar\'e section on the energy surface and the
standard definition of Lyapunov exponents (see \cita{ER}), one would
have found an identically vanishing pairing error. In this respect, it
is interesting to observe that the values and the fluctuations in time
of the pairing error are always much smaller than the fluctuations of
the largest zero Lyapunov exponent $\l_0$ (see Fig. \equ(fig1)(c)),
again indicating a strong cancellation between the exponents.

In the IE case (Fig. \equ(fig1)b and \equ(fig1)d) we observe again a 
$C/t$ behavior for
the two zero Lyapunov exponents, with $C=6.0$ and $C=0$ respectively. 
However, the pairing error does not show any clear behavior and strongly
fluctuates, even after a very long time. Nevertheless, it must be recognized 
that its value is very small and its fluctuation is of the same order of
magnitude as its value. We think that it is very hard to decide 
on the basis of these data if such
a small value is a numerical artifact or a real effect, but we think
that the fact that the fluctuations of the pairing error are of the
same order of magnitude as those of the maximum Lyapunov exponent, suggest 
that, even if pairing is obtained, it could not be SCPR in the strong 
sense of \cita{DM}.

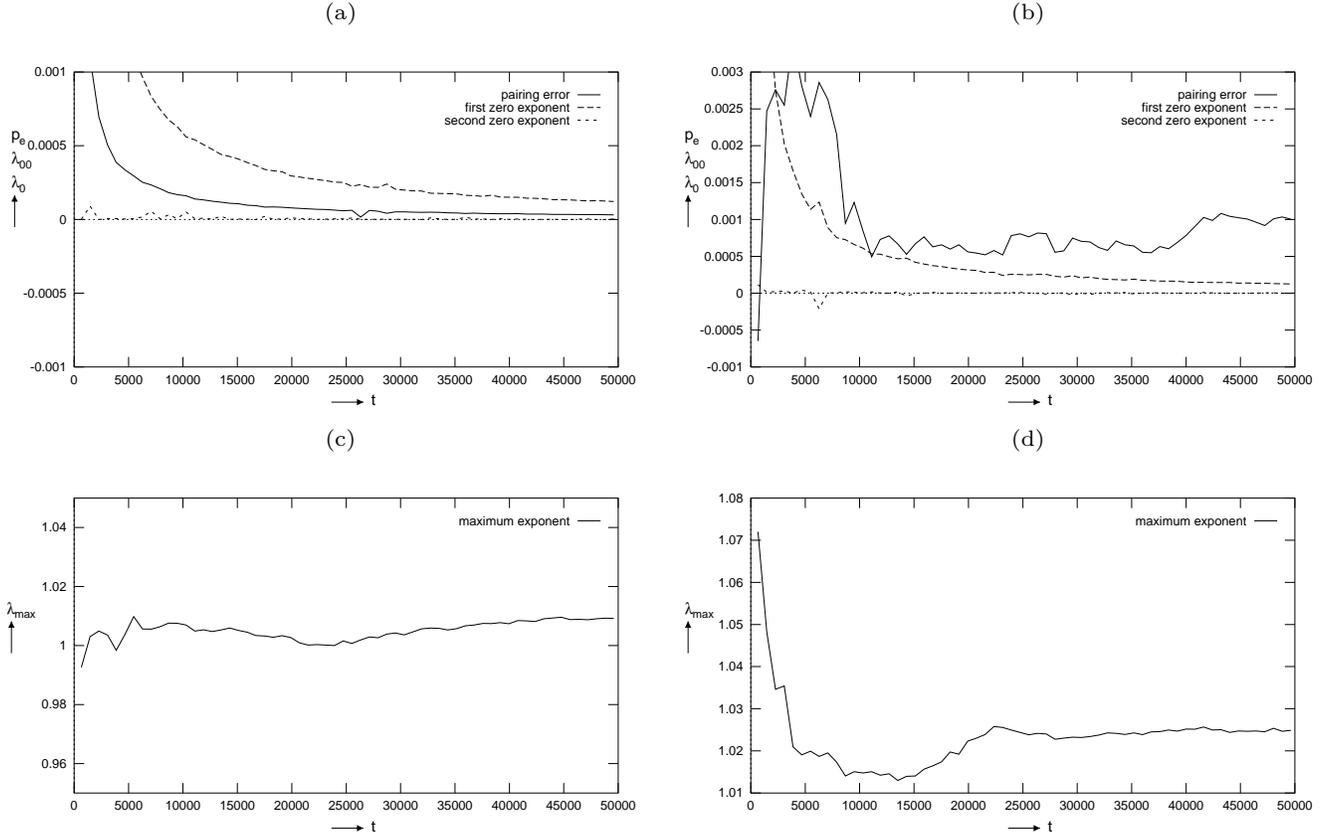
\gnuin isok0 isoe0 l1k1 l1e1 fig1 {Behavior, as functions of time, 
of (a) pairing error, $\l_0$ and $\l_{00}$ as a function of time for
IK Lorentz system (b) same for IE Lorentz system (c) maximum exponent
$\l_{max}$ for IK Lorentz system (d) same for IE Lorentz system }

The results after $50{,}000$ unit of time, 
with the errors computed from the variance of
the last $5{,}000$ time units, are reported in the following table that
summarizes the above discussions. 

\vbox{$$
\halign{\strut\vrule\kern 3mm\hfill#\hfill\kern 3mm&\vrule\kern
3mm\hfill#\hfill\kern 3mm&\vrule\kern 3mm\hfill#\hfill\kern 3mm&\vrule\kern
3mm\hfill#\hfill\kern 3mm&\vrule\kern 3mm\hfill#\hfill\kern 3mm
\vrule\cr
\noalign{\hrule}
&$p_e$&$\l_0$&$\l_{00}$&$\l_{max}$\cr
\noalign{\hrule}
IE&$9\cdot 10^{-4}\pm3\cdot 10^{-4}$&$1\cdot 10^{-4}\pm2\cdot
10^{-5} $&$1\cdot 10^{-6}\pm 9\cdot 10^{-6}$&$ 1.0090\pm 2\cdot 10^{-4}$\cr
\noalign{\hrule}
IK&$4\cdot 10^{-5}\pm1\cdot 10^{-5}$&$1\cdot
10^{-4}\pm1\cdot 10^{-5}$&$2\cdot 10^{-6}\pm3\cdot 10^{-6}$&$1.0248\pm 4\cdot 10^{-4}$\cr
\noalign{\hrule}
}$$
\nobreak
\didascalia{Table 1: Values for pairing error ($p_e$) zero
Lyapunov exponents ($\l_0$ and $\l_{00}$) and maximum Lyapunov
exponent ($\l_{max}$) for IK and IE systems after $50{,}000$ time
units} }

To double check these results we also ran a simulation with
10 different initial conditions, both for the IK and the IE system.
The results are given in Fig. \equ(fig2)(a),(b) where we show 
figures of the behavior of the pairing error $p_e$ for a simulation
with ten initial points, plotting the mean value of $p_e$ over the ten
points, as well as the maximum and the minimum values observed. We
also present the behavior of the maximum Lyapunov exponent in
Fig. \equ(fig2)(c),(d) to compare the error of the pairing error with
that of the maximum exponent.  These results confirm the previous
discussion and the errors evaluated with this method are of the same
order as those previously reported. Namely: for the IK systems the
difference between the maximum and minimum values of the pairing
errors is much smaller than those for the maximum Lyapunov exponent,
while this is not the case for the IE system.  It is interesting to note
that after a long time the pairing errors for all the 10 points of the
IE simulations are positive. Although not conclusive we do think
that this can be considered as strong evidence of non pairing.

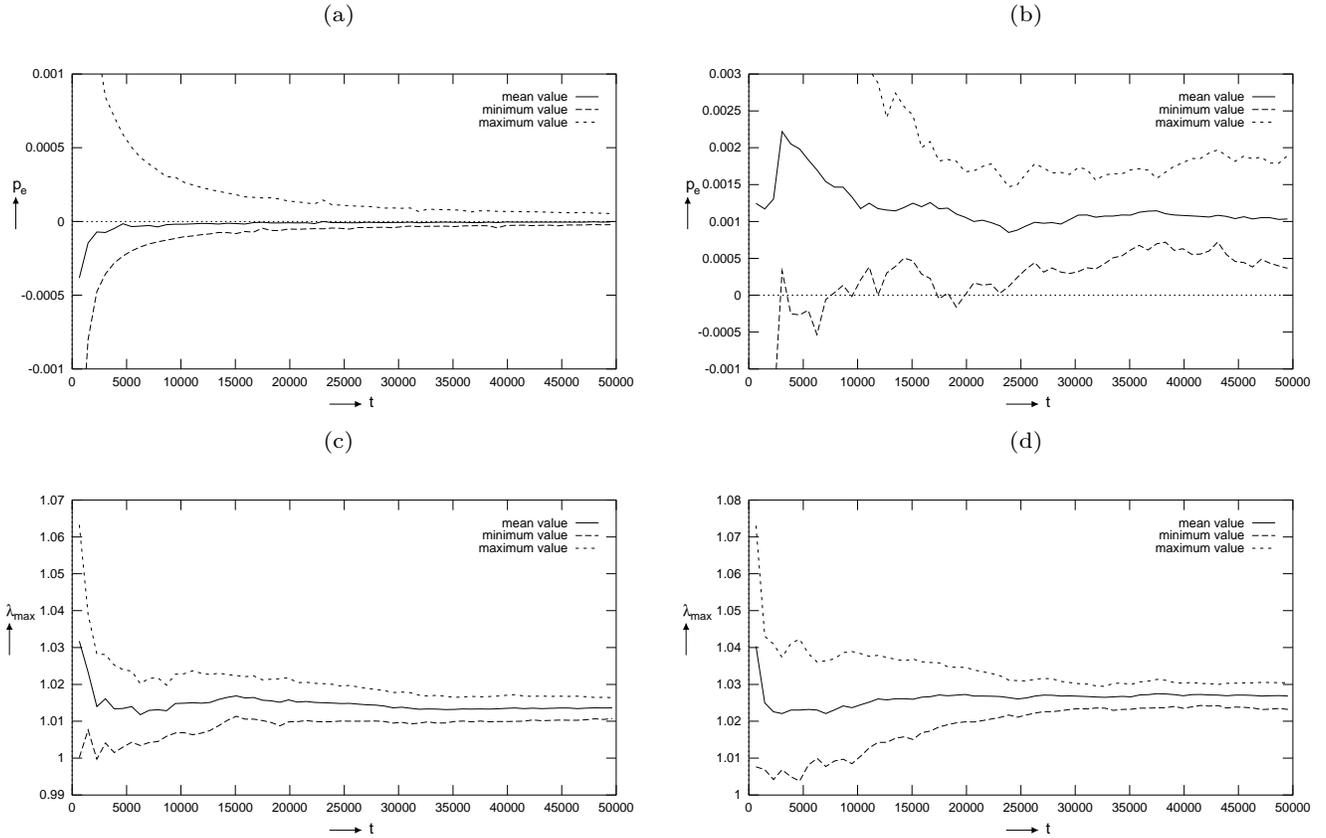
\gnuin isok10 isoe10 l1k10 l1e10 fig2 {Behavior of the (a)  
mean, maximum and minimum values of the pairing error for 10 initial
conditions as a function of time for IK system (b) same for IE system
(c) mean, maximum and minimum values of the maximum Lyapunov exponent
for 10 initial conditions as a function of time for IK system (d) same
for IE system }
\*
\0{B. Generalizing the DM scheme.}
\*
In \cita{DM} Dettmann and Morriss consider the flow $\varphi_t$ on $R^{2d}$
generated by the Hamiltonian IK system \equ(Hami)-\equ(moti). 

The main point in the proof of strong pairing is to split off the two
zero Lyapunov exponent associated with the energy conservation and
flow direction from the other exponents and then consider the pairing
of the other Lyapunov exponents in an appropriate subspace of the
full phase space. To eliminate the exponent associated with the energy
it is enough to restrict the flow to the energy surface. To get rid of
the other exponent we can use a Poincar\'e section like
procedure (see Appendix A2, and figure therein, for a more 
detailed discussion). This means that given a trajectory, 
its initial point ${\bf
x}(0)=({\vec{p}}(0), {\vec{q}}(0))$, a later point ${\bf
x}(t)=({\vec{p}}(t), {\vec{q}}(t))$ and two corresponding planes 
$V_{{\bf x}(0)}$
and $V_{{\bf x}(t)}$ (see Fig. \equ(fig10)), one passing through 
${\bf x}(0)$ and the other through
${\bf x}(t)$, both transversal (i.e. not tangent) to the trajectory,
we can consider the map $\Phi_t$ induced by the flow $\varphi_t$ that
associates to a point ${\bf r}$ on $V_{{\bf x}(0)}$ the point $\Phi_t({\bf r})$
on $V_{{\bf x}(t)}$ which is the first intersection between the trajectory
starting at ${\bf r}$ and the plane $V_{{\bf x}(t)}$. If we properly choose a
plane $V_{{\bf x}(t)}$ for any $t$, the above procedure defines a new flow
$\Phi_t$ whose Lyapunov spectrum is identical to the one of
$\varphi_t$ but does not contain the zero exponents associated with
the flow direction. This is the scheme followed in \cita{DM} where it
is shown that if one chooses as $V_{{\bf x}(t)}$ the plane orthogonal to the
vector $J{\bf e}_0(t)$ with $J=\pmatrix {0&I\cr-I&0\cr}$ the usual
symplectic matrix and ${\bf e}_0(t)=({\vec{p}}(t),0)$ (we will call
this choice of $V_t$ the DM subspace), then the eigenvalues of the
matrix $L^{tr}_tL_t$ are paired, where the tangent flow $L_t=D\Phi_t$
is the differential of $\Phi_t$ (see eq.\equ(diffa)).  From this, the
pairing of the Lyapunov exponents for $\varphi_t$ follows immediately.
Observe that $J{\bf e}_0$ is a vector field defined on the energy
surface with no reference to the particular trajectory. We will call
any vector field ${\bf g}$, like $J{\bf e}_0$ above, such that the
eigenvalues of $L^{tr}_tL_t$, given by the above construction, are
paired an $\alpha$-symplectic field. Observe that $L_t$ satisfies 
an equation of the form:

$$\dot L_t=T({\bf x}(t))L_t\Eq(infi)$$ 
where $T({\bf x})$ is called
the infinitesimal generator of the tangent flow $L_t$.  In \cita{DM}
it is shown that the matrix $L^{tr}_tL_t$ has paired eigenvalue if
$T({\bf x})$ satisfies $JT({\bf x})+T({\bf x})^{tr}J=-\alpha({\bf x})
J$ for a suitable $\alpha({\bf x})$.  If this last condition is
satisfied we call $L_t$ and the dynamics that generate $L_t$
$\alpha$-symplectic and $T({\bf x})$ infinitesimally
$\alpha$-symplectic.

An interesting property of the Lorentz model (with not too high scatterer 
densities, as considered here) is that the particle spends
a significant part of its time in regions where no potential is
present, i.e. between two successive interactions with the scatterers. 
In these regions the IE system is identical to an IK system. It is thus 
evident that if we consider a trajectory for a time small enough that the
particle never interacts with a scatterer, we will find that the local
exponents in the DM subspace are paired. Here by local exponents we
mean those computed as the eigenvalues of $\frac{1}{2t}
\log L^{tr}_tL_t$, where $L_t$ is the differential of the flow (see \cita{ER}).
A natural question is now that if one considers a trajectory which
starts in free space, goes through a collision and then arrives again
in free space, can one find again pairing, using the DM subspace? This
question is interesting because if there is a $\alpha$-symplectic
field for the IE system, this must give pairing for short (never
colliding) trajectories as well as for long trajectories. If no
pairing occurs in the DM subspace for long trajectories with
collisions, this implies that, even if there is a
$\alpha$-symplectic field for the IE system, there must be either
another tranversal symplectic field for the IK system, (at least for
the portion of phase space outside the scatterer potentials, \ie in
free space) or, alternatively, no DM subspace within which pairing
exists for the IE system.

It is not difficult to investigate this question numerically if we
consider only a one collision trajectory and we avoid too singular
collisions. For, in that case we have only to follow the trajectory
for a short time and the diagonalization of the differential of the
flow $L$ can be done with very accurately so that we can achieve a
precision on the local exponents of $10^{-12}$ The result is that for
the IE system after one collision there is no more pairing in the DM
subspace. The interesting point is that the pairing error is of the
same order of magnitude as, \ie consistent with, the one observed
asymptotically in the long time simulations of Fig. \equ(fig1),
discussed sub A. We were able to go through more than one collision
but not through many, because the matrix elements of the tangent flow
grow exponentially. Nevertheless for around three collisions, one sees
that the pairing error in the DM space is always violated by more or
less the same amount, most during collisions and decreasing during
free flight (cf. Fig. \equ(fig3)).

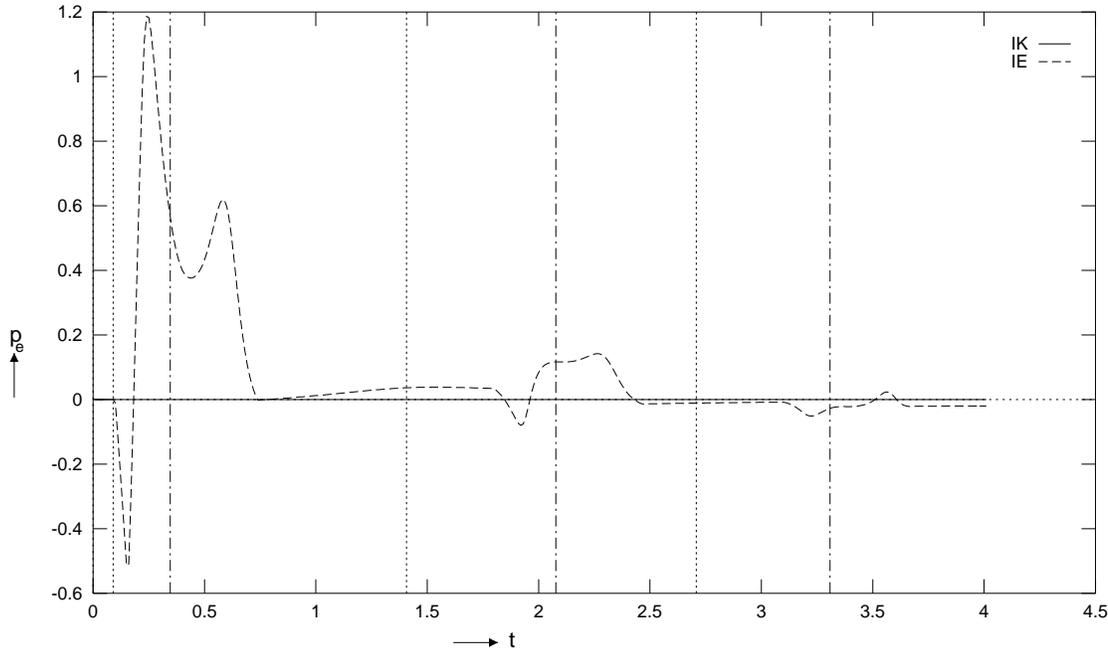
\gnuins anyt fig3 {Behavior of the pairing error as function of time
during a few interaction with the scatteres for the IK (identically 0) and IE system. The time which the particle enters and leaves the potential of a scatterer are indicated by vertical lines of two different dashing (the particle start outside all the potentials).} 

The question of the existence of a possibly different transversal
symplectic field for the IE Lorentz system is nonetheless still open
but this is clearly a question that is not easily investigated
numerically. We chose to look at a similar but in principle simpler 
alternative: given
a point ${\bf x}(0)=({\vec{p}}(0),{\vec{q}}(0))$ and a tangent vector
${\bf w}(0)$, at ${\bf x}(0)$ is it possible to find a vector ${\bf w}(t)$ 
tangent at ${\bf x}(t)$ such
that the linearized flow from the orthogonal to ${\bf w}(0)$ to the
orthogonal to ${\bf w}(t)$ has paired local Lyapunov exponents?  This means
computing the linearized flow for a time $t$ starting from ${\bf x}(0)$ and
then looking for a solution of the equation $p_e({\bf w}(t))=0$ (where we
consider the pairing error as a function of ${\bf w}(t)$). Observe that this
equation depends on four variables, while a unique condition is
imposed so that if a solution exists it will generically not be
unique. Note that the existence of ${\bf w}(0)$ and ${\bf w}(t)$ 
for any initial
point ${\bf x}(0)$ and $t$ does not imply that these vectors generate a
vector field on the energy surface (like is the case in the DM proof,
see Appendix A2). However, this would be enough to prove SCPR, if one also 
had that ${\bf w}(t)$ does not become orthogonal to $\dot{\bf x}(t)$ 

The problem with this computation is that we can handle only short times
$t$ for the same reasons as given above, namely that the 
matrix elements of the tangent flow grow exponentially with the maximum 
Lyapunov exponent while we want to compute a function of them, \ie the minimum 
eigenvalue, which decreases exponentially with the minimum 
Lyapunov exponent. 

In our simulations we followed the trajectory for a time 10 (i.e., for
5 to 7 collisions) computing ${\bf w}(t)$ for $t=n\d t$ with $\d
t=0.1$ and $n=1,2,3,4,...$.  The initial point was chosen outside all
the scattering potentials in such a way that the choice ${\bf
w}(0)=J{\bf e}_0(0)$ was a natural one. To obtain ${\bf w}(t)$, we
started with $JDH_0$ (see \equ(H0),\equ(H00) and \equ(diffa)) looking
for a solution with a steepest-descent method, till the pairing error
changed sign and then using bisection.  Observe that in this setting
the algorithm applied to the IK system would be completely trivial.
We also tried to find other solutions for example using as an initial
attempt for ${\bf w}(t)$ at $t$ the last computed ${\bf w}(t-\d t)$ at $t-\d
t$. We note that this is also a possibly interesting procedure for the
IK case because the result need not be $J{\bf e}_0(t)$, as would be the case for 
\cita{DM}.

In fact, in all cases we were always able to find the vector ${\bf
w}(t)$.  That it was also possible for the IK system to find a ${\bf
w}(t)$ different from $J{\bf e}_0(t)$ which satisfies the pairing rule (at
least for a short time $t$) could lead to a different scheme to prove
CPR for the IK system in a possibly more general fashion than in
\cita{DM}.  Unfortunately we were unable to prove
this and our numerical simulations were too short to give real insight
into the above question.

\newsec{Periodic orbits} 

To obtain a more or less definitive result on the possibility of having
pairing for all time and all initial conditions, we decided to check the
pairing of the Lyapunov exponents on periodic orbits of the single particle
periodic Lorentz gas, considering that periodic orbits solve the problem of
obtaining truly asymptotic Lyapunov exponents,  
since the asymptotic Lyapunov exponents on a periodic orbit are just the
logarithms of the modulus of the eigenvalues of the differential of the
flow after a period, divided by the period.
However, errors are introduced in 
finding periodic orbits numerically. Having a periodic orbit or an orbit
we know is near enough to a periodic orbit, will permit us to obtain a very 
good estimate of the error involved in the computation of the asymptotic
Lyapunov exponents. In fact, if the period of the orbit is not
too long the exponential propagation of errors can be controlled with
an accurate integration scheme (we again used a $4^{th}$ order RK with
adaptive step size with error under $10^{-14}$).   Having to diagonalize a
matrix only once, these operations do not increase the error
significantly for sufficiently short periodic orbits, because, after a 
short time, the matrix elements of the differential of the flow are not 
yet very large. The real problem is to find the best approximate periodic 
orbits numerically and to estimate their deviations from real
periodic orbits. 

To this end we ran a long time single particle trajectory and recorded
the positions of the particle when it entered in the interaction (WCA)
potential of a scatterer (we will register this as a collision). In
this way we get a long list ${\bf x}_n$ of collision points and, fixing a
period $T$, we look for all $m$ such that $|{\bf x}_m-{\bf
x}_{m+T}|<\epsilon$ for some $\epsilon$ ($10^{-2}$ in our
experiments). We then search for a periodic orbit near this orbit using
the Newton-Ramson method. The algorithm stops when the distance
between the initial and the $T$-th collision point is less than
$10^{-12}$.

Having a small distance between the initial and the final point of the
orbit does not automatically ensure that the orbit is near a periodic
one. A good check of this is to see if the application of the
Newton-Ramson method gives a quadratic convergence as should
theoretically be the case$^{\cita{F}}$. This means that iterating the
method, if the error at the $n$-th step is $\epsilon$, it should be
$C\epsilon^2$ at the $(n+1)$-th step, where $C$ is a constant linked
to the first and second derivatives of the flow. Another good check is
to see whether in the computed Lyapunov spectrum there are two 0
exponents (which means that the spectrum of the differential of the
flow must contain two eigenvalues 1 with sufficient precision. Both
these conditions were always fulfilled in our computations: every time
a periodic orbit was found, the algorithm converged quadratically and
there were always two eigenvalues with logarithms of the order of
$10^{-12}$. We think that the error on the evaluated exponents can be
conservatively estimated as less than $10^{-9}$.

We found typically about 200 periodic orbits for periods ranging from 2 to 8.
For all these periods it was possible to find periodic orbits whose exponents 
were not paired by an amount significativelly larger than the expected error.
We report in the following table a number of
relevant quantities we have observed for each period for the two
periodic orbits with the largest pairing errors.

$$\vcenter{\halign{\strut\vrule\kern4mm\hfill#\hfill\kern4mm&
\vrule\kern4mm\hfill#\hfill\kern4mm&
\vrule\kern4mm\hfill#\hfill\kern4mm&
\vrule\kern4mm\hfill#\hfill\kern4mm
\vrule\cr 
\noalign{\hrule}
&pairing level&pairing value&relative pairing error\cr
\noalign{\hrule}
Period 2 &-0.038903& 0.081320& -2.090341 \cr
\noalign{\hrule}
&0.005453& -0.528786& -0.010312 \cr
\noalign{\hrule}
Period 3 &-0.388106& 0.074560& -0.192114 \cr
\noalign{\hrule}
&0.061834& -0.313116& -0.197480 \cr
\noalign{\hrule}
Period 4 &-0.339775& 0.157961& -0.464897 \cr
\noalign{\hrule}
&0.144549& -0.256445& -0.563666 \cr
\noalign{\hrule}
Period 5 &-0.271768& 0.093026& -0.342300 \cr
\noalign{\hrule}
&0.084853& -0.259349& -0.327179 \cr
\noalign{\hrule}
Period 6 &-0.397877& 0.119123& -0.299397 \cr
\noalign{\hrule}
&0.097328& -0.352683& -0.275963 \cr
\noalign{\hrule}
Period 7 &-0.338662& 0.022185& -0.065507 \cr
\noalign{\hrule}
&0.020747& -0.438584& -0.047305 \cr
\noalign{\hrule}
Period 8 &-0.467454& 0.030299& -0.064817 \cr
\noalign{\hrule}
&0.020875& -0.525150& -0.039751 \cr
\noalign{\hrule}
}}$$
\didascalia{Table 2: Pairing value $-\langle\alpha_{IE}\rangle_t$, 
pairing error (eq.\equ(pe)) and relative pairing error (ratio of the
pairing error to the pairing level, i.e. of columns 2 and 1) for each
period of the 2 periodic orbits with the largest pairing errors.}
 
These results provide a numerical proof that SCPR for the one particle
IE Lorentz model cannot be true, so that no proof like that of
\cita{DM} is possible for this system. Nevertheless we note that
this does not rule out the possibility of asymptotic pairing with respect
to an ergodic measure. In fact, periodic orbits are a set of
measure zero so that even if  no periodic orbit has paired
Lyapunov exponents, this does not imply that the asymptotic Lyapunov
exponents 
for $T \rightarrow \infty$ are not paired. 

We note that the behavior of periodic orbits can also lead to a better
understanding of the general behavior of the system.  First, there is 
apparently a large difference between the behavior
of orbits with even and odd periods (even or odd refers here to the number of
collisions on an orbit). In fact, while almost all period 3 orbits have a 
quite large pairing error, those of even period have predominantly a zero
pairing error (i.e. smaller than the precision of the algorithm and
often actually evaluated as zero). To show this effect we made histograms
of the probability $Pr(p_e)$ of finding an orbit with a pairing error $p_e$,
\ie the number of orbits found with that pairing error divided by the total 
number of orbits found. The results are shown in Fig. \equ(fig4).

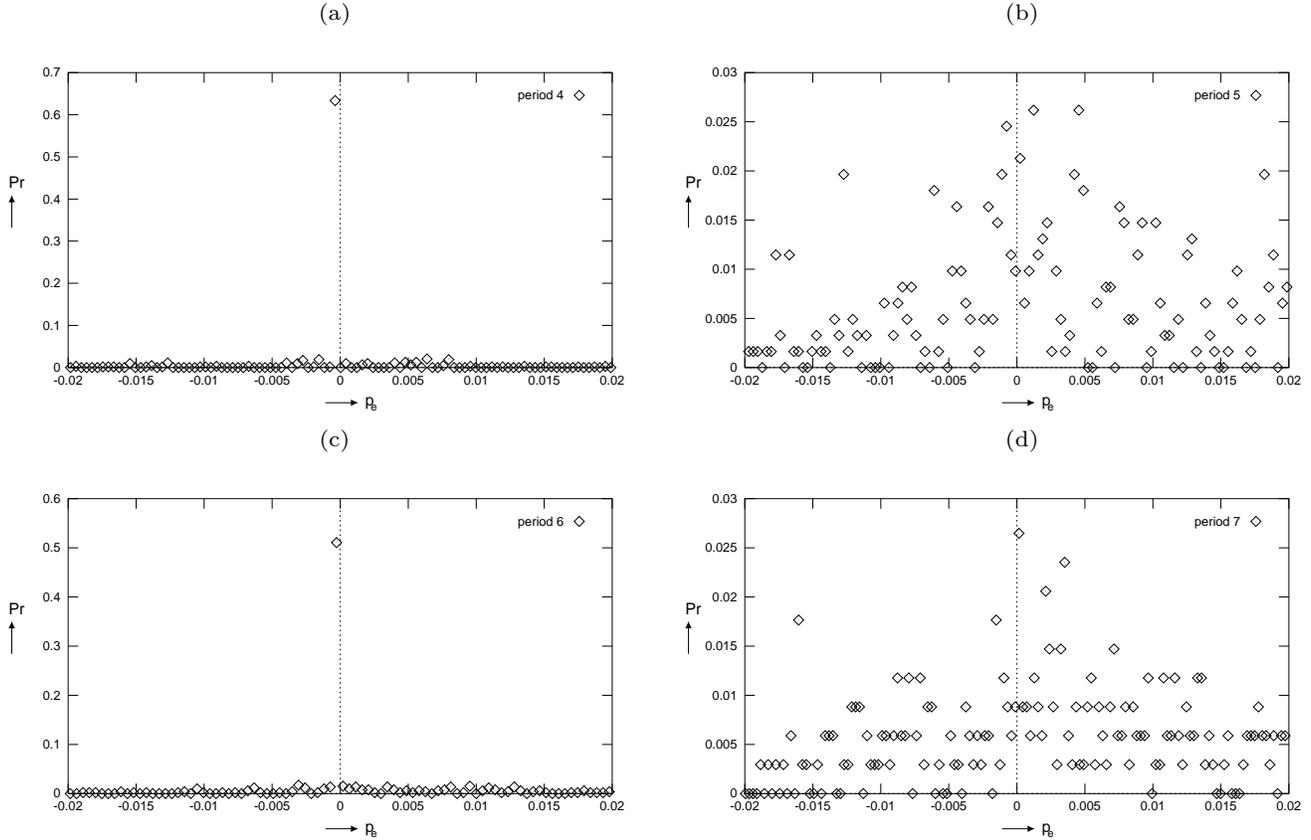
\gnuin p4 p5 p6 p7 fig4 {Distribution of the pairing error for
orbits of period (a) 4 (b) 5 (c) 6 (d) 7. In fig.4(a) the single point
at about 0.65 means that about 65\% of the periodic orbits with period 4
have approximately zero pairing error, while in Fig. 4(b) only about
1\% of the periodic orbits have a pairing error near zero. The 
apparently layered structure of the histograms in Figs 4(b),(d) is 
due to the relatively small intervals chosen in the histograms.}

Secondly, we note that the existence of many even periodic orbits with paired 
Lyapunov exponents can perhaps be connected with some generalization 
of the \cita{DM} proof of pairing.  We hope to come back to this point 
in a future paper.  

\newsec{Many particle systems.} 

In the previous sections we have seen that that no SCPR holds for a
particular Hamiltonian IE system, i.e. the one particle Lorentz model.
In particular, the numerical results on periodic orbits seem to
establish that even asymptotic pairing is violated. In order to
investigate the effect on pairing in the presence of more than one
particle, in fact of many particles, we looked at the same model but
with more than one particle moving among the scatterers.  We
considered color diffusion in an external field, putting no
interaction potential between the particles in order to be able to
make very long and careful simulations, as well as for other reasons
that will become clear in what follows. Observe that when a thermostat
is added to this system of non-interacting particles, the particles
are no more independent, because the thermostat (IK or IE) constraint
introduces an indirect interaction between them.

The equation of motions of this system are again \equ(Hami)-\equ(moti) with 
$N>1$ (we will use $N=2,3,8,16,32$), $\phi^{ext}(\V q)=\sum_{i=1}^N 
(\vec E,\vec q_i)$, using the same notation as in sec. 2 (see comment above 
eq.\equ(WCA1)) and $\phi^{int}(\V q)= \sum_{i=1}^N \phi^{WCA}({\vec q}_i)$ 
with $\phi^{WCA}({\vec q})$ given by eqs.\equ(WCA1)\equ(WCA2). The units are 
fixed as in sec. 2 (see comments after eq.\equ(WCA2)).

In this setting it is possible to consider also a different kind of
thermostat, which we will call an isopeculiar kinetic (IPK) thermostat,
which is, in general, more physical, but has no meaning for a single particle. 
It consists, in constraining only the peculiar (i.e., thermal) kinetic energy, 
i.e the kinetic energy with respect to the center of mass of the moving 
particles. The equations of motion one then obtains are $(i=1,...,N)$:

$$\left\{\eqalign{
\dot{\vec{q}}_i&=\frac{{\vec{p}}_i}{m}\cr
\dot{\vec{p}}_i&={\vec{f}}^{WCA}(\vec q_i)+{\vec{E}}-\alpha_{IPK}
(\V p,\V q)\tilde{\vec{p}}_i\cr}
\right.\Eq(lortN)$$
where ${\vec{f}}^{WCA}(\vec q)=\partial\phi^{WCA}(\vec q)/partial \vec
q$ and
$$\tilde{\vec{p}}_i={\vec{p}}_i- \frac{1}{N}\sum_{j=1}^N {\vec{p}}_j$$
is the peculiar momentum and
$$\alpha_{IPK}(\V p,\V q)= 
\frac{\sum_i( \tilde{\vec{p}}_i,{\vec{f}}^{WCA}(\vec q_i))}
{\sum_i \tilde{\vec{p}}_i^2}\Eq(aIPK)$$

The main reason to look at a many particle system is the idea that the 
CPR could be valid only in the limit that the number of particles
goes to infinity, which would suffice for statistical mechanics. 
There are two kinds of evidences for this point. The
first is based on many numerical simulations, in which the Lyapunov 
spectrum of a many particle system is computed and a good agreement with
the pairing rule is obtained, see \cita{EM}\cita{ECM1}\cita{SE}. However,
considering the small deviations observed in sec. 2 and the periodic
orbit results, it seems doubtful that those simulations can really 
establish pairing definitively.

A second reason arises from an analytic argument that was first presented in
\cita{ECM1}. We give here a more rigorous and extended presentation of that
argument.  

Suppose we have an IE system given by a Hamiltonian of the form
eq.\equ(Hami), equations of motion eq.\equ(moti) with thermostat
eq.\equ(aIE).  In Section 2, sub B, following \cita{DM}, conditions
for pairing of an IK system to hold were tied to finding a proper
subspace.
As was pointed out before the argument for choosing the proper subspace was 
related to the need of taking out of the spectrum the two zero eigenvalues
that are always present, but pair to a different constant 
($\not=0$) than the other Lyapunov exponents. 
However, since we have been unable to find the proper CPR subspace for the
IE system, we were forced (as in the previous section) to proceed in the 
full phase space, replacing,
in a way, the proper subspace by the condition that the zero Lyapunov
exponents $\l_0 \approx \l_{00}$ must be $\approx 0$ in the full 
phase space calculations for CPR to hold.

We can now consider the tangent flow associated with these equations but, 
instead of looking at it in an appropriately chosen transversal subspace,
like we did in sec. 2, we look, in the spirit of the previous section,  at the 
tangent flow in the full phase space. 
Let us call again $\varphi_t({\bf x})$ the flow generated by eq.\equ(moti)
and $L_t=D\varphi_t$ the tangent flow. Then $L_t$ satisfies an equation identical 
to eq.\equ(infi), with ${\bf x}=(\V p,\V q)$:

$$\dot L_t=T({\bf x}(t))L_t\Eq(sym)$$ 
(we use here the same convenient notation in full phase space as we did 
in the DM subspace 
of sec.2 for slightly different objects, 
because we think that no confusion can arise).  The matrix $T({\bf x})$, the
infinitesimal generator of the tangent flow, has the simple form:

$$T({\bf x})=T_H({\bf x})-\alpha_{IE}({\bf x}) I_{\V{p}} - \V p\otimes\V\nabla \alpha_{IE}({\bf x})
\Eq(TIE)$$
where $T_H$ is the infinitesimal generator of the tangent flow for the
Hamiltonian system, i.e. $T_H=J\V \partial^2 H$ with $J$ the usual
symplectic matrix and $\V \partial^2 H$ the matrix of the second
derivatives of $H$ with respect to all momenta and positions. Moreover
$I_{\V p}=\pmatrix{I&0\cr0&0}$, i.e. the identity on the $\V p$
subspace and $\otimes$ represents the tensor (or dyadic) product.  It is
easy to see that if we neglect the term $\V p\otimes\V\nabla 
\alpha_{IE}({\bf x})$
the matrix $\check T({\bf x})=T_H({\bf x})-\alpha_{IE}({\bf x}) I_{\V p}$ 
satisfies the relation:

$$J\check T({\bf x})+\check T^{tr}({\bf x})J=-\alpha_{IE}({\bf x})
J\Eq(TIEa)$$ 
where $\check T^{tr}({\bf x})$ is the transpose of
$\check T({\bf x})$.  This follows from the fact that for a
Hamiltonian system $J T_H({\bf x})+ T_H({\bf x})^{tr}J=0$ and $J I_{\V
p}+I_{\V p}^{tr}J=J$.  This, in turn, implies that if we consider the
characteristic exponents$^{\cita{ER}}$ associated with the matrix
$\check T({\bf x})$, which we will call the mutilated spectrum of the
system, they will be paired. This follows directly, for example, from
the last part of \cita{DM}, or from \cita{ECM1} and the pairing level
will be exactly equal to $-<\alpha_{IE}>_t$.  Moreover for these
exponents a SCPR holds.  Note that the mutilated spectrum does not
necessarily have two zero exponents.  This creates a first difference
between the (true) Lyapunov exponents spectrum and the (mutilated)
characteristic exponents spectrum, but we expect that the two zero
Lyapunov exponents associated with $T({\bf x})$ will be converted into
a couple of paired mutilated characteristic exponents associated with
$\tilde T({\bf x})$, one of which is numerically found to be zero,
while the other one equals twice the pairing level $-<\alpha_{IE}>_t$
(see Fig. \equ(fig7) and discussion below it).  The critical question is
whether the term $\V p\otimes\V\nabla \alpha({\bf x})$ can be
neglected. Observe that
$$\eqalign{{\partial \alpha_{IE}({\bf x})\over\partial \vec p_i}=&
{f_i^{ext}(\V q)\over \sum_j \vec p_j^2}+ 
\alpha_{IE}({\bf x}){{\vec{p_i}}\over \sum_j\vec p_i^2}\cr
{\partial \alpha_{IE}({\bf x})\over\partial \vec q_i}=&0\cr}\Eq(diff)
$$
If we go to the thermodynamic limit, keeping the energy per particle 
fixed, we will have that, in the mean $\sum_j \vec p_j^2 \simeq N||\vec p_i||$,
so that we can say that $\V\nabla \alpha_{IE}$ is of the order of $1/N$.  
Although this makes the above procedure more reasonable, it is mathematically 
not justified. A similar argument holds for IPK	thermostatted systems. 
In fact we have in that case:

$$T^{IPK}({\bf x})=T_H({\bf x})-\alpha_{IPK}({\bf x})\tilde I_{\tilde{p}}-
\tilde{\V{p}}\otimes\V\nabla\alpha_{IPK}({\bf x})\Eq(TIPK)$$
where $\tilde I_{\tilde{\V p}}=I_{{\V p}} + O(1/N)$ and $\V\nabla
\alpha_{IPK}({\bf x})$ is given by slightly more complicated expressions
than eq.\equ(diff), but can still be considered as being $O(1/N)$.

The $1/N$-argument has three major problems:

\item{a)} It is mathematically  speaking not always true that 
${\partial \alpha_{IE}\over\partial \vec p_i}=O(1/N)$. 
For, in the IE case it can 
happen that $\sum_j\vec p_i^2$ is very small or even zero, if all the energy 
of the system is potential energy, so that the system has no kinetic energy. 
This implies 
then that the first term in eq.\equ(diff) can be very large. Of course, it will
nevertheless be true that for most parts of phase space $\sum_j\vec p_j^2$ is 
$O(N)$ and the volume of the region where it is small will go to zero as the 
number of particle goes to infinity.
This effect is not present in the IPK system, because in the denominator of
$\alpha_{IPK}$ the constrained quantity itself then appears.

\item{2)} If we consider the two equations
$$\dot L_t=T_tL_t\qquad\qquad\dot {\check L_t}=\check T_t\check L_t$$
where $\check T=T+O(1/N)$, it is not generically true that $\check L_t=
L_t+O(1/N)$ uniformly in $t$. In fact we will typically have that 
$$\check L_t=L_t+f(t)O(1/N)\Eq(app)$$
for some $f(t)$. The point is then to control the growth of $f(t)$, but this 
is, generically, very hard.

\item{3)} Even assuming that we could prove \equ(app), it is easy to
give an example of matrices  of the form $\check L_N=L_N+(1/N) P_N$ in which
the spectrum of $\check L_N$ is completely different from that of $L_N$.
Let $L_N$ a Jordan block with
eigenvalue $a$. That is $(L_N)_{i,i}=a,(L_N)_{i,i+1}=1$, while all
other matrix elements are zero. Let $(P_N)_{N,1}=1$ while all other
matrix elements of $P_N$ are zero, i.e., 
$$L_N=\pmatrix{a&1&0&\cdots&0&0\cr0&a&1&\cdots&0&0\cr0&0&a&\cdots&0&0\cr
\vdots&\vdots&\vdots&\ddots&\vdots&\vdots\cr0&0&0&\cdots&a&1\cr
0&0&0&\cdots&0&a}\qquad
P_N=\pmatrix{0&0&\cdots&0\cr0&0&\cdots&0\cr\vdots&\vdots&\ddots&\vdots\cr
1&0&\cdots&0}
$$
Using the properties of triangular matrices it is easy to see that 
the spectrum of $L_N$ consists then only of the point
$a$ counted with multiplicity $N$, while the eigenvalues $\l_i$ of $\check L_N$
are given by $\l_i=N^{-\frac{1}{N}}u_{i,N}$ where $u_{i,N}$ are the $N$-th 
root of the unity, \ie $u_{i,N}^N=1$. This implies that the spectrum of 
$\check L_N$ converge to the  circle in the complex plane of radius 1 
with center at $a$, when $N$ goes to infinity.
%
%
Moreover, if we replace the perturbation $O(N^{-1})$ by a perturbation
$O(f(N)^{-1})$ we need essentially that
$\lim_{N\to\infty}{f(N)\over\exp(N)}=\infty$ for the spectrum of 
$\check L_N$ to coincide with that of $L_N$, i.e., any
perturbation of the form $N^{-p}$ for a fixed $p$ will give the same
result. 
This is clearly connected with the existence of many
degenerate eigenvalues, but it is not clear that a hypothesis of
non degeneracy would be enough to give convergence of the spectrum of 
$\check L_N$ to that of $L_N$.

Since we were not able to decide analytically the above points, which are
well known difficulties if one tries to prove continuity of
characteristic exponents with respect to perturbations, we checked the
behavior of the Lyapunov exponents as a function of $N$ numerically.
We did simulations for $N=2, 4, 8, 16$ and $32$ particles. In this case it
became not feasible to run simulations for 10 different initial
conditions, because the 16 and 32 particle simulations already needed
two weeks of CPU time. So we computed errors by only looking at the
fluctuations in the last part of the trajectory (see sect. 2) and we
always ran analogous simulations for the IK model to check the
precision of the program.  Moreover the lengths of the simulations
were necessarily shorter than those used for the single particle
system.  Observe that, for the IE system, because of the absence of
interactions among the particles, if we neglect the term containing
the derivatives of $\alpha$, the matrix $\check T$ splits into a block
matrix consisting on $N$ blocks of dimension $6\times6$ on the
principal diagonal of the matrix. This, assuming ergodicity, implies
that the mutilated spectrum for this system consists of only 6
different characteristic exponents, each of which is $N$ times
degenerate.  This allows a very easy check on the convergence of the
spectra of the characteristic exponents associated with $\check T$, as
$N$ goes to infinity.

Fig. \equ(fig5) shows the behavior of the pairing error as a function
of time, each graph reporting the observed pairing error for the IK,
IE, IPK system at a fixed number of particles. To make the comparison
more informative we decided to plot the behavior of the ratio $\P_e$
between the pairing error $P_e$ and $-<\alpha>_t$, i.e.  the value at
which the exponents would pair if pairing were to occur. Observe that
in this situation the number of pairs is no more 2, as it was in the
one particle system.  Therefore, we adopted as a definition of pairing
error the variance of the values observed, i.e. calling
$m_i=\lambda_i+\lambda_{6N-2-i}$ we set
$$P_e=\frac{1}{N}\sum_im_i^2-\left(\frac{1}{N}\sum_im_i\right)^2\Eq(Pe)$$.

Observe that the behavior in time of the pairing error for the IE and
IPK systems becomes more and more regular and similar to that of the IK
system as the number of particles increases. In section 2 we saw that
the smooth behavior of the pairing error of the form $C/t$ 
for the IK system can be
deduced from the existence of strong cancellations leading to
a proof of SCPR. This suggests that a form of SCPR could
hold when the number of particle goes to infinity or, at least, that
the local Lyapunov exponents in a proper subspace will pair faster 
and faster in time.

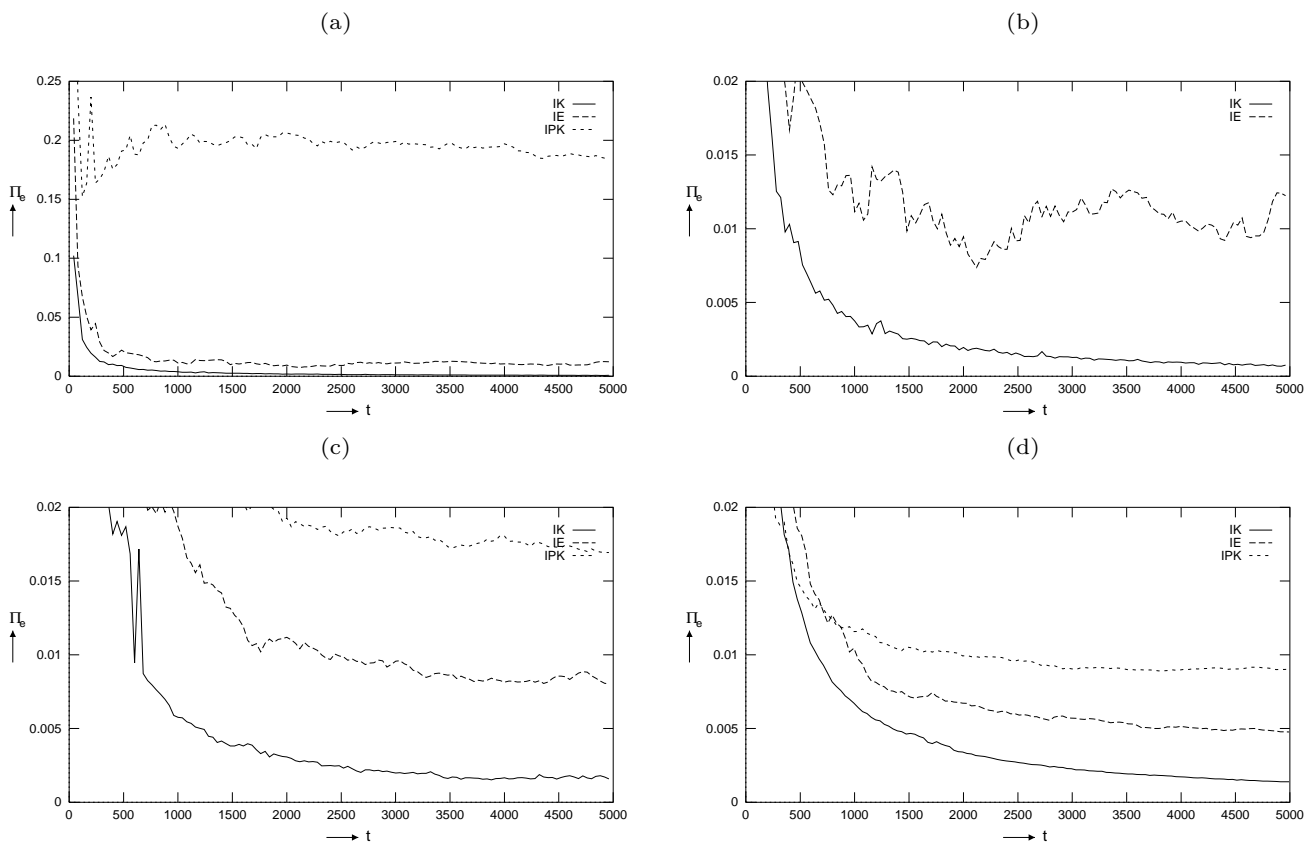
\gnuin comp2 comp2a comp8 comp32 fig5 {Behavior of ratio $\P_e$ of the
pairing error $P_e$ and $- < \alpha>_t$ as a function of time, 
for the IK, IE, IPK systems with (a) 2 particles; 
(b) IK, IE of (a) on a finer scale; (c) 8 particles; (d) 32 particles.}

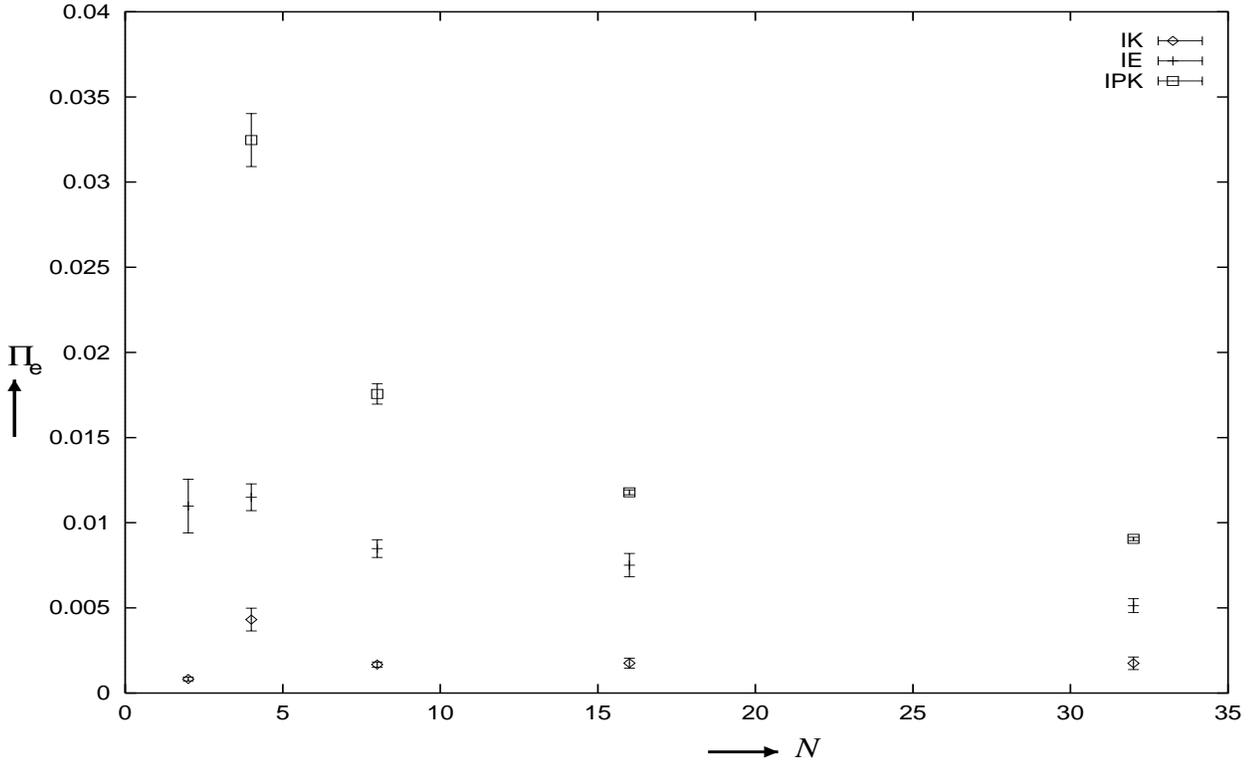
\gnuins compf fig6 {Ratio $\P_e$ of the
pairing error $P_e$ and $- < \alpha>_t$  as a function of the number of
particles for the IK, IE, IPK systems. The IK values provide an estimate
of the computer program accuracy, since they should vanish.}

In Fig. \equ(fig6) we report the values after $5{,}000$ time units for $\P_e$
of the three systems as a function of the number of particles $N$.

We see that the pairing error for the IK, IE and the IPK systems
is a decreasing function of $N$. Moreover the final value of the last
two systems gets closer to the final value obtained for the IK system,
where we know that the exponents should be paired. Thus it appears 
that our simulations are
consistent with pairing for the IE system in the thermodynamic limit.

Finally, as noted before, the characteristic exponents associated with
the matrix $\check T$ have a particularly simple structure and are
quite easy to evaluate. So we checked to what extent the Lyapunov
spectrum converges to this simple spectrum. Fig. \equ(fig7) reports
the results for the IE system while Fig. \equ(fig8) reports those for
the IPK system. Note that in these figures only $3N-1$ pairs of
exponents are shown, because one pair can be easily deduced, \ie
it is a pair of zero exponents for the Lyapunov spectrum, while 
it is identical to
the $(3N-1)$-th for the mutilated spectrum.  Moreover, one would expect
in general that the IE system and the IK system would become more and
more similar as the number of particles goes to infinity.  This is
indeed seen in Fig. \equ(fig9).  Note that this argument does not
apply to the IPK case. In fact this thermostat has a different and
more physical nature in that the velocity of the center of mass is not
constrained and only the energy with respect to the center of mass is
fixed, so that we cannot expect any equivalence between the IE or IK
systems and the IPK system.

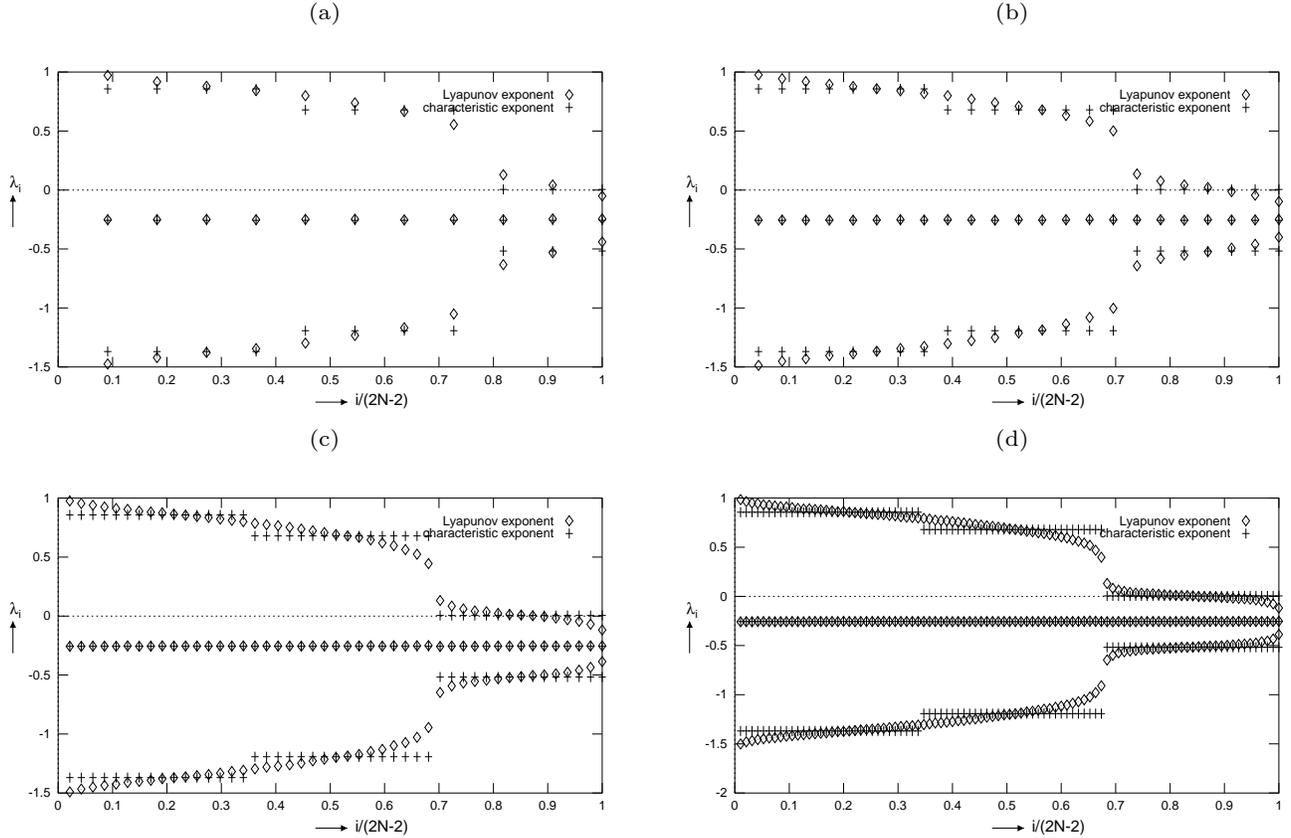
\gnuin tl4e tl8e tl16e  tl32e fig7 {Comparison of the
Lyapunov spectrum $\l_i$ of an IE system with the spectrum of characteristic
exponents associated with  $\check T$ for (a) 4 particles; 
(b) 8 particles; (c) 16 particles; (d) 32 particles.}

Both Figs 7 and 8 show a weak similarity of the spectrum associated with the
matrix $\check T$ and the real Lyapunov spectrum.  However, the two
spectra do not seem to converge to each other for $N \rightarrow \infty$.  
The spectra of both models could in principle both split for $N \rightarrow 
\infty$ into six straight segments, an indication of which is visible in 
Fig. \equ(fig7), at two thirds of the Lyapunov spectrum.  Since it is
rather unlikely that the spectrum associated with $T$ will become
discontinuous when $N \rightarrow \infty$, it seems much more likely
that the discontinuous spectrum associated with $\check T$ will become
continuous, if direct interactions between the particles are introduced.
It is interesting to note that the two systems appear to pair to the
same value and it is almost impossible to see from the figures whether
the exponents are exactly paired or not. The observed pairing
error of the Lyapunov spectrum is so small that any deviation of
pairing could well be due to numerical errors or to too short a
simulation time. The same cannot be said of the spectrum themselves.

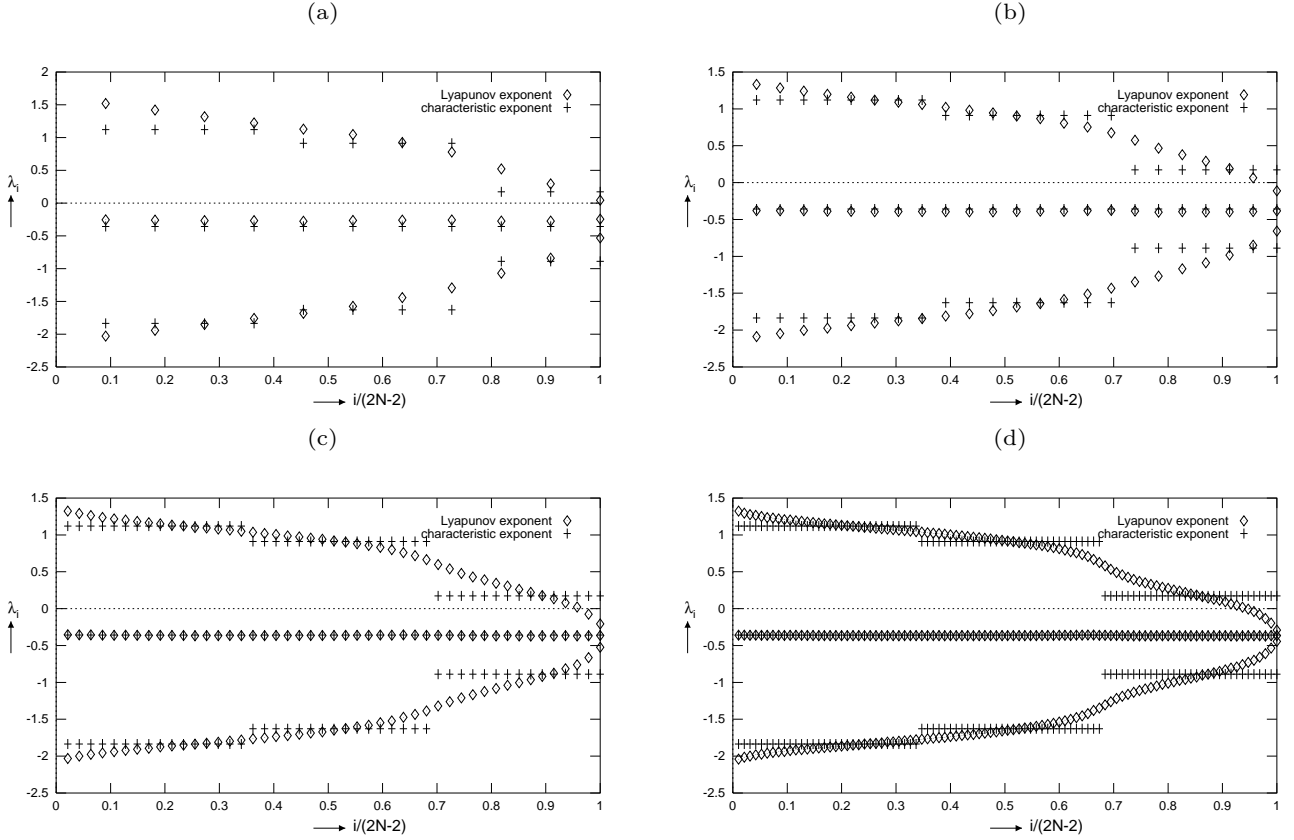
\gnuin tl4i tl8i tl16i  tl32i fig8 {Comparison of the
Lyapunov spectrum $\l_i$ of an IPK system with the spectrum of characteristic
exponents associated with $\check T$ for (a) 4 particles (b) 8 particles (c) 16
particles (d) 32 particles.}

Finally we discuss Fig. 9. A generalization of the idea of the
equivalence of ensembles for large $N$, discussed formally in a
different context by a number of authors$^{\cita{ESG}}$, makes it
quite reasonable to surmise that the IE and IK systems should become
equivalent for sufficiently large $N$. This is quite consistent with
our computations. We see that the mean phase space contraction $- <
\alpha >_t$ is the same for the two thermostatted system, while the
two spectra agree very well for the larger exponents, differing only
slightly for the smaller exponents, in the latter parts of the
spectra.

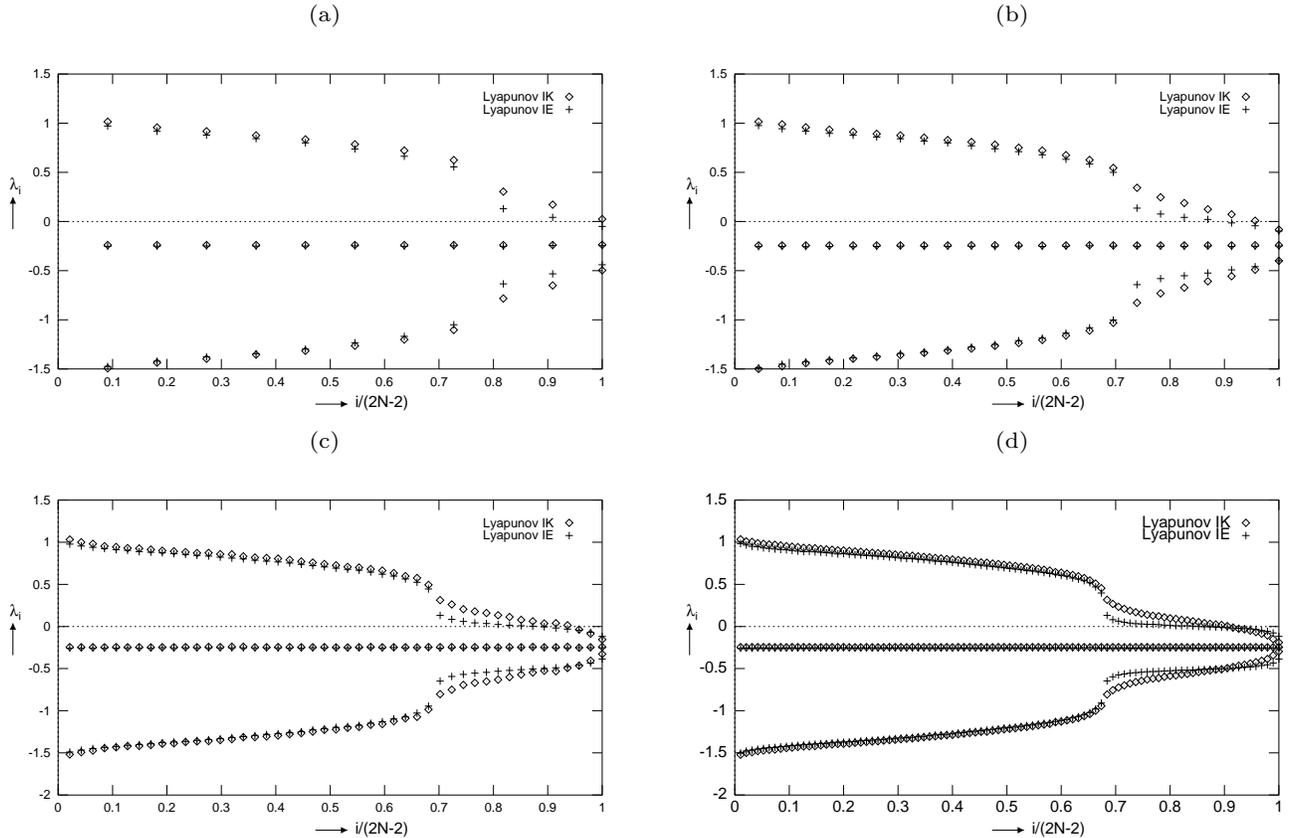
\gnuin cm4 cm8 cm16  cm32 fig9 {Comparison of the
Lyapunov spectrum $\l_i$ of an IE system with the characteristic 
spectrum of an IK 
system associated with $\check T$ for (a) 4 particles; (b) 8 particles; 
(c) 16 particles; (d) 32 particles.}

\newsec{Discussion}

\item{1.}One of the main results of this paper is the numerical 
proof of the non validity of SCPR for a one particle Lorentz gas. This
is obtained through a careful evaluation of periodic orbits, keeping
meticulously track of the computational errors. In fact, Lyapunov
exponents for periodic orbits can be computed with very high accuracy
if the periodic orbit is not too long. The existence of periodic
orbits that do not satisfy the strong pairing rule shows that SCPR is
not valid. We also note that periodic orbits form a set of measure
zero. This implies that no conclusion as to asymptotic pairing of
non-periodic trajectories can be drawn from the above
observations. Nevertheless the violation of pairing in periodic orbits
is consistent with the results of the very long simulations for
non-periodic trajectories of Section 2, used to compute asymptotic
Lyapunov exponents. Also these simulations showed deviations from
pairing for the IE systems, but the deviations were too small to
identify clearly a real effect, because of the numerical errors
always associated with this kind of simulations.

\item{2.}We have also tried to generalize the DM scheme for proving 
strong pairing.  This lead to a reformulation of the proofs in
\cita{DM}, \cita{LiWo} and \cita{D} by one of us (C.P.).  With reference to the
discussion and notation of sec. 2 sub B, we call a vector field ${\bf
g}$ for the flow $\varphi_t$ defined on the energy surface
$\alpha$-symplectic if the eigenvalues of $L^{tr}_t({\bf x})L_t({\bf
x})$ are paired, where $L_t$ is the linearized flow for the trajectory
starting at ${\bf x}$.  Then it was possible to give a necessary and
sufficient condition on ${\bf g}$ to be $\alpha$-symplectic
directly in terms of the equations of motion . More precisely, calling
$V_{\bf x}$ the plane orthogonal to ${\bf g}({\bf x})$ in the point ${\bf
x}$ and $T({\bf x})={d \over dt}L_t({\bf x})\bigg|_{t=0}$, this
implies that there must exist a function $\alpha({\bf x})$, such that
the matrix $M({\bf x})=JT({\bf x})+T^{tr}({\bf x})J$ satisfies:
$$({\bf w},M({\bf x}){\bf w}')=\alpha({\bf x})({\bf w},{\bf
w}')\Eq(Ch)$$ 
for any vectors ${\bf w},{\bf w}'\in V_{\bf x}$. For more
details we refer to Appendix A2.

\item{3.}To test the idea that for a large number of particles (i.e. in 
the thermodynamic limit) the IK and IE thermostats should give
identical results we have done a number of simulations on a number of
many particle Lorentz gases. The necessity to evaluate the full
Lyapunov spectrum with high accuracy made it necessary to use high
order integration methods leading to the impossibility of using a
number of particles in excess of 32. In this setting we have
considered also a different kind of thermostat (IPK) in which only the
kinetic energy respect to the center of mass of the moving particles
is constrined. The results of the simulations seem to be in good
agreement with the hypothesis that pairing should hold (possibly in a
strong sense) for IE and IPK systems in the thermodynamic limit.
This hypothesis was first used heuristcally in \cita{ECM1}. However,
as shown in section 4, it cannot easily be converted into a proof of CPR.
Nevertheless, the numerical results obtained here are consistent with this 
hypothesis, thus suggesting an interesting property of large thermostatted 
Hamiltonian systems. 

\*
\0{\bf Acknowledgement}
\*
Two of us (FB and EGDC) gratefully acknowledge financial support by
grant DE-FG02-88-ER13847 of the US Department of Energy, as well as
from ESI in Vienna. One of us (F.B.) is endebted to C. Liverani and
M. Wojtkowski for helpful discussions at ESI. C.P. expresses his thanks to 
the Rockefeller University for support during 1996--1997.
We also acknowledge helpful discussions with C. Dettmann.
\appendices

\newapp{Hamiltonian equivalence} 

In \cita{DM2} it was shown that an IK system for a given value of the energy
can be mapped into a Hamiltonian system on an energy surface by
a proper time rescaling and redefinition of the momenta. More generally
it was shown in \cita{LiWo} that any $\alpha$-symplectic dynamics
can be mapped into a symplectic one ($\alpha = 0$) one. 
In \cita{DM2} there is no
explicit derivation of the proper rescaling, but it is quite easy to derive 
this starting from a Hamiltonian system. We note that if such a
transformation to a Hamiltonian system can be found for an IE system,
this will quite easily lead to a proof of SCPR. Moreover we believe
that, if SCPR holds for an IE system, some kind of mapping into a
Hamiltonian system should be possible$^{\cita{F}}$. To be general enough we
consider a Hamiltonian $H$ of the form eq.\equ(Hami) with equations of motion
given by eq.\equ(moti)\equ(aIE).

We now try to find a function $t(\lambda)$ of some new time variable
$\l$such that $\frac{1}{2}\sum_i
{\vec{q}}_i\,{}'(t(\l))^2+\phi^{int}(\V q)={\cal E}$ for some constant
${\cal E}$ representing the new fixed internal energy. Here the prime 
represents the derivative respect to $\lambda$, \ie
${\vec{q}}_i\,{}'(t(\l))=d{\vec{q}}_i(t(\l))/d\l$.  This leads, if we
restrict ourself to the $H=E$ energy surface, to:
$$t'(\l)^2={{\cal E}-\phi^{int}(\V q)\over E-\phi^{int}(\V q)-\phi^{ext}(\V q)}
\Eq(lt)$$

Defining $\V \pi=t'(\l)\V p(t(\l))$ we get 
$$\left\{\eqalign{
{\vec{q}}_i\,{}'&={\vec{\pi}}_i\cr
{\vec{\pi}}_i\,{}'&={\vec{F}}_i^{ext}(\V q)+{\vec{f}}_i^{int}(\V q)-
\alpha(\V \pi,\V q)\vec\pi_i\cr
\alpha(\V \pi,\V q)&={\sum_i {\vec{\pi}}'_i {\vec{F}}^{ext}_i(\V q)
\over\sum_i {\vec{p}}_i^2}}\right.\Eq(risc)$$

where 
$${\vec{F}}_i^{ext}(\V q)={({\cal E}-\phi^{int}(\V q))({\vec{f}}_i^{ext}(\V q)
+{\vec{f}}_i^{int}(\V q))
\over E-\phi^{int}(\V q)-\phi^{ext}(\V q)}-{\vec{f}}_i^{int}(\V q)\Eq(iso)$$

It is now easy to see that if we set $\phi^{int}(\V q)=0$ we obtain:
$${\vec{F}}_i^{ext}(\V q)={\cal E}{{\vec{f}}_i^{ext}(\V q)\over
E-\phi^{ext}(\V q)} = {\cal E} \vec\nabla_i\log (E-\phi^{ext}(\V
q))\Eq(isoDM)$$ which is exactly the inverse transformation of the one
in \cita{DM2}.

The system \equ(iso) looks very much like an IE system. The only
problem is that the force ${\vec{F}}_i^{ext}(\V q)$ is not a gradient
unless $\phi^{ext}(\V q)$ is a multiple of $\phi^{int}(\V q)$
\cita{DD}. This shows that it is, in general, not possible to map an
IE system into a Hamiltonian system, at least using the simple
mappings discussed here.

\newapp{A geometric condition for strong pairing}

In general, for the thermostatted systems we are considering, the
Lyapunov spectrum contains two zero exponents, one due to the presence
of a constant of the motion ( kinetic energy for IK, internal energy
for IE and peculiar kinetic energy for IPK) and the other due to the
motion in the time direction. To see whether the other Lyapunov
exponent pair to a non zero value, one has to conceive a well defined
procedure to split off the two zero exponents from the spectrum. This
is done in \cita{DM} by a Poincar\'e section like procedure. We want
to give here a general version of this procedure, leading to an
explicit algebraic condition on the differential of the flow for
having strongly paired Lyapunov exponents.

As a general setting we consider the phase space of our system as
being a $2n-1$ sub-manifold $\Sigma$ of $R^{2n}$ (the full $(\V p,\V
q)$ phase space) for some integer $n$ defined by a particular level
set of the constant of the motion (\ie the set of points on which the
constant of motiom takes a given value). The flow itself, denoted by
$\varphi_t$, is generated by a vector field $\V {\bf v}$, so that
$$\dot \varphi_t({\bf x})=\V{\bf v}(\varphi_t({\bf x}))\Eq(flow)$$
for ${\bf x}\in\Sigma$.

We will also consider in $R^{2n}$ the standard symplectic matrix
$J=\pmatrix{0&I\cr -I&0\cr}$, already defined in sect. 2,
where $I$ is the $n\times n$ identity matrix. It seems natural
to introduce $J$ considering that, both in the Hamiltonian and in the IK case
pairing comes from symplectic properties of the flow $\varphi_t$. In
general, we can consider $\Sigma$ as a $2n-1$ submanifold of some more
general symplectic manifold ${\cal M}$ but we will consider only the case
${\cal M}=R^{2n}$ with the symplectic structure given by $J$.

We now apply this to the pairing property.  The Lyapunov spectrum of
the flow $\varphi_t$, with $\varphi_t$ defined only on $\Sigma$, does
not contain the zero exponent due to the imposed constant of the
motion. To eliminate also the second zero exponent in the time
direction we proceed as follows, following \cita{DM}. At any point
${\bf x}\in\Sigma$ we consider an hyperplane $V_{\bf x}$ of $R^{2n}$
at ${\bf x}$ and uniformely transversal to ${\bf v}_{\bf x}$, \ie the
angle between $V_{\bf x}$ and ${\bf v}_{\bf x}$ is everywhere bounded
away from 0. The intersection between $V_{\bf x}$ and $\Sigma$ defines
a $(2n-2)$-dimensional submanifold $\Omega_{\bf x}$ whose tangent
space $E_{\bf x}$ is just the intersection between the tangent space
of $\Sigma$ at ${\bf x}$, $T_{\bf x}\Sigma$, and $V_{\bf x}$ (here and
in what follows we will identify $R^{2n}$ with its tangent space at
any point, so that $T_{\bf x}V_{\bf x}=V_{\bf x}$). Observe that the
family $V_{\bf x}$ can be defined as the family of hyperplanes
orthogonal to a vector field ${\bf g}_{\bf x}$ on $\Sigma$ ( in
general ${\bf g}_{\bf x}$ need not be in $T_{\bf x}\Sigma$, it can
be in $R^{2n}$, but in what follows we will always consider ${\bf
g}_{\bf x}\in T_{\bf x}\Sigma$). The family $V_{\bf x}$ will allow us
to define a new flow, $\Phi_t$, whose Lyapunov spectrum will not
contain the second zero exponent as well.

Given now a point ${\bf x}(0)\in\Sigma$ and its time evolved point ${\bf
x}(t)=\varphi_t({\bf x}(0))$ after a given time $t$, we can define a
new map $\Phi_t$ from a neighborhood of ${\bf x}(0)$ in $\Omega_{{\bf
x}(0)}$ to $\Omega_{{\bf x}(t)}$ as follows. Given a point ${\bf x}'(0)\in
\Omega_{{\bf x}(0)}$ close enough to ${\bf x}(0)$, we consider the time
$\tau({\bf x}'(0),t)$ at wich the trajectory starting from ${\bf x}'(0)$
intersects $\Omega_{{\bf x}(t)}$. Here and in the following we suppress
the dependence on ${\bf x}(0)$. Setting then $\Phi_t({\bf
x}'(0))=\varphi_{\tau({\bf x}'(0),t)}({\bf x}'(0))$ it is clear that
$\Phi_t({\bf x})$ can be considered as a new flow and that its
Lyapunov spectrum, \ie the eigenvalues of
$\Lambda=\lim_{t\to\i}\frac{1}{2t}\log D\Phi_t^{tr} D\Phi_t$, are
identical to those of $\varphi_t$ at ${\bf x}(0)$, except that they do not
contain the two zero exponents. Here $D\Phi_t: V_{{\bf x}(0)}\to V_{{\bf
x}(t)}$ is the differential of $\Phi_t$ and is given in local
coordinate by:

$$\bigl(D\Phi_t\bigr)_{i,j}=\frac{\partial (\Phi_t)_i}{\partial
{\bf x}_j}\Eq(diffa)$$
\eqfig{350pt}{250pt}{
\ins{144pt}{137pt}{${\bf x}(0)$}
\ins{130pt}{112pt}{${\bf x}'(0)$}
\ins{256pt}{144pt}{${\bf x}'(t)$}
\ins{218pt}{136pt}{${\bf x}'(\t)$}
\ins{215pt}{157pt}{${\bf x}(t)$}
\ins{90pt}{47pt}{$\Sigma$}
\ins{125pt}{50pt}{$\O_{{\bf x}(0)}$}
\ins{225pt}{55pt}{$\O_{{\bf x}(t)}$}
\ins{165pt}{10pt}{$V_{{\bf x}(0)}$}
\ins{265pt}{15pt}{$V_{{\bf x}(t)}$}
\ins{97pt}{225pt}{$E_{{\bf x}(0)}$}
\ins{167pt}{225pt}{$E_{{\bf x}(t)}$}
\ins{90pt}{144pt}{${\bf g}_{{\bf x}(0)}$}
\ins{141pt}{158pt}{${\bf n}_{{\bf x}(0)}$}
\ins{182pt}{154pt}{${\bf g}_{{\bf x}(t)}$}
\ins{235pt}{174pt}{${\bf n}_{{\bf x}(t)}$}
}{fig3}{\geq(fig10)}
\*
\didascalia{Fig. \equ(fig10): 
Schematic representation of the construction of the
flow $\Phi_t$. Here ${\bf x}(t)=\varphi_t({\bf x}(0))$, 
${\bf x}'(t)=\varphi_t({\bf x}'(0))$ and 
${\bf x}'(\tau)=\Phi_t({\bf x}'(0))\in \Omega_{{\bf x}(t)}$. The other 
symbols are explained in the text.}
\global\setbox150\vbox{\unvbox150 \*\*\0 Fig. \equ(fig10): 
Schematic representation of 
the construction of the flow $\Phi_t$. 
Here ${\bf x}(t)=\varphi_t({\bf x}(0))$, 
${\bf x}'(t)=\varphi_t({\bf x}'(0))$ and 
${\bf x}'(\tau)=\Phi_t({\bf x}'(0))\in \Omega_{{\bf x}(t)}$. The other 
symbols are explained in the text.}
\*
Oberve that, calling $L_t= D\Phi_t$, we have $\dot L_t=T({\bf x}(t))L_t$
where $T({\bf x})$ is a matrix uniquely defined at every point ${\bf
x}$ of $\Sigma$, independently of ${\bf x}(0)$, contrary to $\Phi_t$
and $L_t$. 
In fact we have:

$$T({\bf x})_{i,j}=\frac{d}{dt}\bigl(D\Phi_t({\bf x})\bigr)_{i,j}
=\frac{d}{dt}\biggl(\frac{(\partial\varphi_\tau({\bf x}))_i}{\partial
\tau}\frac{\partial\tau({\bf x},t)}
{\partial {\bf x}_j}+\frac{(\partial\varphi_\tau({\bf x}))_i}{\partial
{\bf x}_j}\biggr)\Eq(T)$$
so that, for ${\bf w}\in E_{\bf x}$ one has, after straightforward
algebraic manipulations:

$$T({\bf x}){\bf w}=\biggl[\frac{d}{dt}\bigl(D\tau\bigr){\bf w}\biggr]
{\bf v}_{\bf x}+
\biggl[\bigl(D\tau\bigr){\bf w}\biggr]\frac{d}{dt}{\bf v}_{\bf x}+
\frac{d}{dt}\bigl(D\varphi_t){\bf w}\Eq(TT)$$
where $D\tau$ is the differential of $\tau$ with respect to ${\bf x}$.

Noting that $\frac{d}{dt}\bigl(D\varphi_t\bigr)=
\bigl(D{\bf v}\bigr)\bigl(D\varphi_t\bigr)$ we obtain:

$$T({\bf x}){\bf w}=\biggl[\biggl(\frac{\partial D\tau}{\partial t}
{\bf w}\biggr)\biggr]{\bf v}+\bigl(D\V{\bf v}\bigr)w\Eq(TTT)$$
It is now easy to show that $\frac{\partial D\tau}{\partial t}=
{\bf b}+\tilde {\bf b}$ with 
$${\bf b}=\frac{\dot {\bf g}+\bigl(D{\bf v}\bigr)^{tr}{\bf g}}{({\bf v},
{\bf g})}\Eq(b)$$ 
and $\tilde {\bf b}||{\bf g}$. 
Moreover the component of the vector
${\bf b}$ along $V_{\bf x}$ is unique. Thus eq.\equ(TTT) can be
written as:

$$T{\bf w}=\bigl(D{\bf v}\bigr) {\bf w}+({\bf b},{\bf w}){\bf v}\Eq(tlast)$$
where ${\bf w}\in E_{\bf x}$ and ${\bf b}$ is given by eq.\equ(b).

It is known, see \cita{DM} for a simple proof, that if 
for any ${\bf w},{\bf w}'\in E_{{\bf x}_0}$

$$(L_t{\bf w},JL_t{\bf w}')=\mu_t({\bf w},J{\bf w}')\Eq(asym)$$
where $\mu_t=\exp(\int_0^t \alpha({\bf x}(\tau))d\tau)$ for some
function $\alpha({\bf x})$ defined on $\Sigma$, then if $\lambda$ is in
the spectrum of $\Lambda_t=L_t^{tr}L_t$, so is $\mu^2/\lambda$. 
This immediately implies that if
$\lim_{t\to\i}\frac{1}{2t}\log\Lambda_t$ exists then its eigenvalues are
strongly paired in the sense of sec. 1. Moreover, for any ergodic measure for
$\varphi_t$ the Lyapunov exponents of $\varphi_t$ are paired.  Observe
that in order give meaning to eq.\equ(asym) we need that $V_{\bf x}$
is a symplectic plane, \ie $JV_{\bf x}=V_{\bf x}$. To achieve this we
must set ${\bf g}=J{\bf n}$ where ${\bf n}$ is the normal vector 
to $\Sigma$ at ${\bf x}$. 
We will say that $L_t$ is $\alpha$-symplectic if
eq.\equ(asym) holds.  Moreover we will say that the flow $\varphi_t$
is $\alpha$-symplectic if for any ${\bf x}_0$ and for any $t$ the
matrix $L_t$ is $\alpha$-symplectic.
We can now state the result.
\*
\\{\bf Theorem \teq(as)}{\it: The flow $\varphi_t$ is $\alpha$-symplectic
if and only if for any ${\bf x}\in\Sigma$ and for all $w,w'\in E_{\bf x}$ the skew
symmetric matrix $M=JT+T^{tr}J$ satisfies:

$$(w,Mw')=\alpha(w,Jw')\Eq(th1)$$
where $\alpha:\Sigma\to R$ is a suitable function and $T$ is given by
eq.\equ(tlast).}
\*
Eq. \equ(th1) follows immediately by differentiating eq.\equ(asym) with respect
to $t$. 

Theorem \equ(as) provides a well defined algebraic condition that can
be easily checked in cases of interest. In fact, straightforward
computation shows that it is indeed satisfied in the case of the IK
dynamics studied by Dettmann and Morris in \cita{DM} as well as in 
the case of constant $\alpha$, studied by Dressler in \cita{D}.

\biblio
\rife{BGG}{BGG}{F. Bonetto, G. Gallavotti, P.L. Garrido, ``Chaotic Principle: an experimantal test'', Physica D {\bf 105}, 226-252 (1997).}

\rife{BGGS}{BGGS}{G. Benettin, L. Galgani, A. Giorgilli,J.M. Strelcyn,``Tous les nombres charact\'eri\-sti\-ques de Lyapunov sont effectivement calculable'', C. R. Acad. Sci. Paris, Ser A--B {\bf 286} A431--A433 (1978).}

\rife{Bo}{Bo}{R. Bowen, {\it Equilibrium states and the ergodic theory of Anosov diffeomorphism}, Lecture Notes in Mathematics {\bf 470}, Springer, Berlin (1975).}

\rife{CR}{CR}{E.G.D. Cohen, L. Rondoni: ``Note on phase space contraction and entropy production for thermostatted systems'', archived in cond-mat@xxx.lanl.gov \#9712213.}

\rife{CELS}{CELS}{N.I. Chernov, G.L. Eyink, J.L. Lebowitz,
Ya.G. Sinai, ``Steady state electric conduction in the periodic Lorentz gas'', 
Comm. Math. Phys. {\bf 154}, 569--601 (1993).}

\rife{D}{D}{U. Dressler, ``Symmetry properties of the Lyapunov spectra of a class of dissipative dynamical systems with viscous damping'', Phys. Rev. A {\bf 38}, 2103-2109 (1988).}

\rife{DM}{DM}{C.P. Dettmann, G.P. Morriss, ``Proof of Lyapunov exponent pairing for system at constant kinetic energy'', Phys. Rev E {\bf 53}, 5545-5548 (1996).}

\rife{DM2}{DM$_2$}{C.P. Dettmann, G.P. Morriss, ``Hamiltonian formulation of the Gaussian Isokinetic Thermostat'', Phys. Rev. E {\bf 54}, 2495-2500 (1996).}

\rife{ECM1}{ECM$_1$}{ D.J. Evans, E.G.D. Cohen, G.P. Morris, ``Viscosity of a simple fluid from its maximal Lyapunov exponents'', Phys. Rev. A {\bf 42} 5590--5597 (1990).}

\rife{EM}{EM}{D.J. Evans, G.P. Morris: {\it Statistical Mechanics of nonequilibrium fluids}, Accademic Press (1990).}

\rife{ER}{ER}{J.M. Eckmann, D. Ruelle,``Ergodic Theory of Chaos and Strange Attractors'',Rev. Mod. Phys. {\bf 57} 617--656 (1985).}

\rife{ESG}{ESG}{D.J. Evans, S. Sarman:``Equivalence of thermostatted nonlinear responses'', Phys. Rev. E {\bf 48}, 65-70 (1993); G. Gallavotti ``Equivalence of dynamical ensamble and Navier-Stokes equations'', Phys. Rev. Lett. {\bf 223}, 91-95 (1996); ibid. ``Dynamical Ensemble equivalence in Statistical Mechanics'', Physica D, in print.}

\rife{F}{F}{M.J. Feigenbaum, Private communication.}

\rife{GR}{GR}{G. Gallavotti: ``Chaotic hypothesis: Onsager reciprocity and fluctuation dissipation theorem'', J. Stat. Phys. {\bf 84}, 899-926, (1996); ibid., ``Extension of Onsager's reciprocity to large field and the chaotic hypothesis'', Phys. Rev. Lett. {\bf 77}, 4334-4337, (1996); G. Gallavotti, D. Ruelle: ``SRB states and nonequilibrium statistical mechanics close to equilibrium'', in mp-arch@math.utexas.edu \# 96-645, in print on Comm. Math. Phys..}

\rife{LiWo}{LiWo}{M Wojtkoski, C. Liverani, ``Conformally symplectic dynamics and symmetry of Lyapunov spectrum'',Comm. Math. Phys. (1997) in print (Jan. 30).}

\rife{DD}{DD}{The same observation was made independently by C. Dettmann.}

\rife{PH}{PH}{H.A. Posch, W.G. Hoover: ``Lyapunov instability of dense Lenard-Jones fluids'', Phys. Rev. A. {\bf 38}, 473--482 (1988).}

\rife{PO}{PO}{H.A. Posch: ``Lyapunov instability of the boundary-driven Chernov-Lebowitz model for stationary shear flow'', J. Stat. Phys. {\bf 88} 825--842 (1997).}

\rife{Ru}{Ru}{D. Ruelle: ``Positivity of entropy production in nonequilibrium statistical mechanics'', in mp-arch@math.utexas.edu \# 96-166.}

\rife{SE}{SE}{D.J. Searles, D.J. Evans, D.J. Isbister, ``The conjugate pairing rule for non-Ha\-mil\-to\-ni\-an systems'', Chaos, Focus issue on ``Chaos and Irreversibility'', (1998).}

\rife{Si}{Si}{Ya.G. Sinai,``Hyperbolicity and stochasticity of dynamical systems'', Math. Phys. Rev. {\bf 2} 53--115 (1981).}

\rife{Z}{Z}{R. Zwanzig:``Time-Correlation Functions and Transport Coefficient in Statistical Mechanics'', Ann. Rev. Phys. Chem. {\bf 16}, 67-102 (1965).}

\ciao